\documentclass[aps,pra,showpacs,superscriptaddress,floatfix]{revtex4}
\usepackage{amssymb}
\usepackage{amsmath}
\usepackage{graphicx}

\begin{document}

\title{The quantum Levy walk}
\author{Manuel O. C\'{a}ceres$^{\ast }$ and Marco Nizama}
\affiliation{Centro Atomico Bariloche, CNEA, and Instituto Balseiro, Universidad Nacional
de Cuyo, Av. E. Bustillo Km 9.5, (8400) Bariloche, Argentina\\
}
\altaffiliation{Senior member of CONICET}
\date{\today }

\begin{abstract}
We introduce the quantum Levy walk to study transport and decoherence in a
quantum random model. We have derived from second order perturbation theory
the quantum master equation for a \textit{Levy-like particle }that moves
along a lattice through hopping scale-free while interacting with a thermal
bath of oscillators. The general evolution of the quantum Levy particle has
been solved for different preparations of the system. We examine the
evolution of the quantum purity, the localized correlation, and the
probability to be in a lattice site, all them leading to important
conclusions concerning quantum irreversibility and decoherence features. We
prove that the quantum thermal mean-square displacement is finite under a
constraint that is different when compared to the classical Weierstrass
random walk. We prove that when the mean-square displacement is infinite the
density of state has a complex null-set inside the Brillouin zone. We show
the existence of a critical behavior in the continuous eigenenergy which is
related to its non-differentiability and self-affine characteristics. In
general our approach allows to study analytically quantum fluctuations and
decoherence in a long-range hopping model.


\end{abstract}

\maketitle

\section{Introduction}

\bigskip

Random walks have been studied for a long time now in order to obtain
information about the effect of dimensionality, symmetry, and topological
structures on the general properties of classical transport phenomena. Via
the Central Limit Theorem the Gaussian distribution plays a fundamental role
for all random walks (dynamic semigroup) with finite mean square
displacement per step \cite{kampen,libro}. Levy flights are classical
Markovian random walks in a continuous space with infinite mean square
displacement per step (thus Levy flights are not in the attractor of the
Gaussian distribution). The interesting point about these walks is that the
set of points visited by these Levy flights has a self-similar clustering
property, so a fractal dimension has been extensively discussed and in
particular magnificently illustrated by Mandelbrot \cite{Mandelbrot}. A
Weierstrass random walk is a discrete (lattice) version of Levy flights
which was introduced by Schlesinger et al. \cite{HSM81}. This transport
model (also called Levy walk) shows the fractal nature of the random walk
trajectory, and several mathematical concepts such as non-differentiable
function, lacunary Taylor series, fractals, renormalization group
transformation, etc., have been related to this Weierstrass random walk and
mapped to important characteristics of transport phenomena \cite{MSh94}.

Recently it has been reported that power-law discrete-stable (step)
distributions can be obtained as the stationary state of immigration Markov
processes \cite{HJM-JPA-L745}, showing that the statistics of the steps of
the particle is inherited from the $m-$tuplets rates (mesoscopic)
fluctuations appearing in the master equation \cite{HJM-JPA-36}. To tackle
this kind of program from a non-classical point of view is much more harder
because in quantum mechanics it is necessary to include the cause of damping
and noise explicitly, thus apart from the Hamiltonian of the system of
interest $\mathcal{S}$ one has to include a thermal bath $\mathcal{B}$ an
interaction between both \cite{kampen,vK95}. The program of the present work
is not in the direction of reference \cite{HJM-JPA-36}, rather our purpose
here can be presented in two fold: first to get a quantum semigroup
description of a mechanical particle that moves doing scale-free hopping
along a lattice of sites, and second to study the quantum decoherence
phenomena in a long-range dissipative model. In order to achieve this
program we will adopt the simplest quantum mechanical assumptions to get an
analytical model.

To gain insight into the relationship between dimensionality, topological
structures and disorder, many quantum models of transport phenomena have
been introduced \cite{Economou}. Among them the \textit{tight-binding}
approximation for a quantum particle over a regular structure with
nearest-neighbor (NN)\ interactions, is a simple description which is
equivalent to the quantum random walk (QRW), a quantum particle that moves
along a lattice of sites doing NN steps while interacting with a bath \cite%
{vK95,MOC-CH97}. One of the most interesting facts that distinguishes
quantum mechanics from classical mechanics is the coherent superposition of
distinct physical states. Many of the non-intuitive aspects of the quantum
theory of matter can be traced to the coherent superposition feature. Two
important question are:

$\bullet $ How does the coherent superposition operate in the presence of
dissipation?

$\bullet $ How does a long-range interaction characterize quantum
decoherence?

These subjects have been important issues of research since the pioneer
works of Feynmann and Vernon \cite{Feynmann1,Feynmann2}, Caldeira and
Leggett \cite{caldeira} among others, see for example the references cited
in: \cite{kampen,plenio,blum,spohn,QN}.

The study of a quantum walk subjected to different sources of decoherence is
an active topic that has been considered by several authors, in particular,
by their interest in understanding Laser cooling experiments \cite%
{Bouchaud03}, modeling Blinking Statistics \cite{Margolin}, and also in
doing quantum simulations \cite{Romanelli}. It should be noted that the most
usual Laser cooling scheme is based on the idea that the microscopic quantum
description of subrecoil cooling can be replaced by a study of a related
random (anomalous) walk in momentum space, this is the point where the
concept of Levy waiting-time, for the statistical description of the elapsed
time between walks, appears. The definition of our Quantum Levy Walk (QLW)
is base in a long-range jump model without introducing any waiting-time
statistics. The physical motivation of using a Levy-like probability for the
jump was inspired on recent experiments showing long-range interaction as in
Rydberg gases \cite{Lukin,Ates,Anderson}. We end this paragraph noting that
our QLW is quite different from the quantum walk with Levy waiting-time.

In this paper we introduce a \textit{scale free} open quantum model, i.e., a
quantum mechanical particle that moves along a lattice through hopping scale
free while interacting with a thermal phonon bath. We have chosen the
interaction Hamiltonian with the bath in such a way that it produces a
long-range superposition of vector states. We highlight some of the issues
of interpretation of the coherent superposition by tackling a soluble
long-range hopping model. The asymptotic long-time regime of the quantum
purity is characterized by a long-tail with a non-trivial exponent that
depends on the Weierstrass parameters of the Hamiltonian. A long-time
coherent behavior for the localized correlation function is also explained
in terms of the present scale free hopping model.

In appendix A we present some formal aspects of quantum dynamic semigroups,
and we revisit the second order approximation obtained for an open quantum
system weakly coupled to the environment \cite%
{plenio,blum,spohn,alicki,kossa,lindblad,davies,QN,kampen}. Thus we
emphasize some general conditions on the system of interest, the environment
and the interaction Hamiltonian to obtain a true semigroup. In the core of
the paper we present our open quantum model: \textit{The Quantum Levy Walk},
which is a generalization of the QRW Hamiltonian \cite{vK95}. Then we obtain
the evolution equation for the reduced density matrix under the Markovian
approximation \cite{jpa,ukranian}, its solution, and solve analytically some
correlation functions associated to the coherent superposition feature.

\section{The quantum Levy model}

For open quantum systems, the Markovian description of the dynamics is based
on the concept of quantum dynamic semigroups \cite{spohn,alicki}. Only with
these semigroups are the properties of the reduced density matrix of the
quantum system of interest, preserved during the whole time evolution
(positivity, trace and hermiticity) \cite{kossa,lindblad,davies}. From a
microscopic description considering the total Hamiltonian of the system of
interest and the environment, it is possible to derive a picture involving
the quantum dynamic semigroups.

One of the pioneers work in obtaining the QRW model from first principles
can be found in van Kampen's paper \cite{vK95}, where it is shown that
tracing out the baths variables a bona-fide semigroup is obtained. Here we
will do something similar, but introducing a scale free Hamiltonian, for the
free particle, and generalizing the interaction with the phonon thermal
bath. Let the system $\mathcal{S}$ be a free particle that can reside in any
lattice site $l=0,\pm 1,\pm 2\cdots ,$ the dynamics of the system $\mathcal{S%
}$ will be described by the Hamiltonian%
\begin{equation}
H_{S}=\Omega \left( 1-\frac{a+a^{\dag }}{2}\right) ,  \label{H}
\end{equation}%
where the shift operators $a,a^{\dag }$ act on the orthonormal set of
Wannier basis $\left\vert l\right\rangle $ that spans the Hilbert space $%
\mathbf{H}$. A general form for these shift operators can be expressed as:%
\begin{eqnarray}
a &=&\sum\limits_{n=0}^{\infty }\sum\limits_{l\in \mathcal{Z}}f(\epsilon
_{n})\left\vert l-\epsilon _{n}\right\rangle \left\langle l\right\vert \
,\qquad \epsilon _{n}\in \mathcal{Z},  \label{a-general} \\
a^{\dag } &=&\sum\limits_{n=0}^{\infty }\sum\limits_{l\in \mathcal{Z}%
}f(\epsilon _{n})\left\vert l+\epsilon _{n}\right\rangle \left\langle
l\right\vert ,  \label{a-general2}
\end{eqnarray}%
with $f(\epsilon _{n})\geq 0$ and fulfilling normalization to one, to help
its physical interpretation.

A particular choice of $f(\epsilon _{n})$ will describe a tight-binding-like
Hamiltonian ranging form NN to long-range hopping, thus the set $a,a^{\dag }$
will be the fundamental operators to modelling the interaction Hamiltonian
between system $\mathcal{S}$ and the bath (a thermal set of oscillators),
see appendix A. A classical one dimensional random walk is defined in terms
of the probability for a particle to make a step of a given length to the
left or to the right. Quantum random walk Hamiltonians are described instead
in terms of probabilities amplitudes (here the shift operators $a$ and $%
a^{\dag }$). Related NN\ discrete-time models are (coined) quantum random
walks \cite{Aharonov}. Also the NN random walk Hamiltonian (in a ring) has
been used to study transport in a quantum trapping model \cite{Blumen07}.
Quantum random walk, is the counterpart of a classical random walk for
particles which cannot be precisely localized due to quantum uncertainties.
When $f(\epsilon _{n})=0,\ \forall \epsilon _{n}\neq \epsilon _{0}$ with
lattice parameter $\epsilon _{0}=1$ we reobtain the usual NN random walk in
the line \cite{vK95}. When $f(\epsilon _{n})\propto n^{-1-\alpha }$, $%
\epsilon _{n}=n$, $\alpha >0$, $n=1,2,3\cdots $ we get the Gillis and Weiss
lattice walk model \cite{Gillis}, etc.

If $f(\epsilon _{n})$ is characterized by a power-law probability we obtain
a Levy-like jumping walk, this type of quantum walk has been previously
reported in order to study long-range interaction in a non-dissipative model 
\cite{Blumen08a}, as well as in a trapping transport model \cite{Blumen08}.
Here we propose to study a jumping model characterized by the Weierstrass
probability, this class of lacunary long-range probability has been widely
study in classical transport phenomena, nevertheless up to our knowledge
nothing has been reported in the context of quantum transport. Two important
consequence appear from the definitions (\ref{a-general}) and (\ref%
{a-general2}): using that $\epsilon _{n}$ belongs to the Wannier index we
can see that $a$ and $a^{\dag }$ are diagonal in the Fourier basis, this
fact leads to the conclusion that also $H_{S}$ will be diagonal, and also it
can be proved that $a$ and $a^{\dag }$ commute. These results will be shown
in the Appendix B for the particular Weierstrass jumping probability.

We define the Weierstrass shift operators in the form:%
\begin{eqnarray}
a &=&\frac{\mathcal{A-}1}{\mathcal{A}}\sum\limits_{n=0}^{\infty
}\sum\limits_{l=-\infty }^{\infty }\frac{1}{\mathcal{A}^{n}}\left\vert
l-b^{n}\right\rangle \left\langle l\right\vert \qquad ;\ \mathcal{A}>1,b(%
\mbox{integer})\geq 1,  \label{Shift} \\
a^{\dag } &=&\frac{\mathcal{A-}1}{\mathcal{A}}\sum\limits_{n=0}^{\infty
}\sum\limits_{l=-\infty }^{\infty }\frac{1}{\mathcal{A}^{n}}\left\vert
l+b^{n}\right\rangle \left\langle l\right\vert \qquad ;\ \mathcal{A}>1,b(%
\mbox{integer})\geq 1.  \notag
\end{eqnarray}%
This means that the application in the Weierstrass shift operators to any
vector $\left\vert l_{1}\right\rangle $ produces a linear combination in the
Wannier basis. This linear combination is scale free and has a clustering
structure \cite{slesinger}. This clustering is characterized --in average--
by the probability $\propto 1/\mathcal{A}^{n}$ to have a projection on a
Wannier vector distant $b^{n}$ from $\left\vert l_{1}\right\rangle $. For
example%
\begin{equation*}
a^{\dag }\left\vert l_{1}\right\rangle =\frac{\mathcal{A-}1}{\mathcal{A}}%
\sum\limits_{n=0}^{\infty }\frac{1}{\mathcal{A}^{n}}\left\vert
l_{1}+b^{n}\right\rangle .
\end{equation*}

Interestingly as we mention before, 
\begin{equation*}
\lbrack a,a^{\dag }]=0;\ \forall \mathcal{A}>1,\forall b(\mbox{integer})\geq
1,
\end{equation*}%
this and other important results are shown in Appendix B.

We noted that only in the case $b=1$ the operators $a$ and $a^{\dag }$ are 
\textit{truly} translation operators, in the sense that the product of
successive translations is equivalent to one resultant translation, i.e.,
for example the application of $n$ translations gives:%
\begin{equation*}
a^{\dag }\cdots a^{\dag }\left\vert l\right\rangle =\left\vert
l+n\right\rangle \qquad \text{only if\qquad }b=1.
\end{equation*}%
Therefore it is simple to see that taking $b=1$ in (\ref{Shift}) we reobtain
the QRW model \cite{vK95,MOC-CH97}.

In general for $b>1$ the Hamiltonian $H_{S}$ describes a scale free
tight-binding-like Hamiltonian (the QLW model), i.e., a quantum free
particle in a lattice that moves, making hopping like a Levy walk, while
interacting with a bath (in the present paper we work in a discrete one
dimensional infinite Hilbert space). The eigenfunctions of $H_{S}$ are
denoted by the kets $\left\vert k\right\rangle $ (with $-\pi <k<\pi $) and
are given by the Fourier transform of the Wannier vectors:%
\begin{eqnarray*}
\left\vert k\right\rangle &=&\frac{1}{\sqrt{2\pi }}\sum\limits_{l=-\infty
}^{\infty }e^{ikl}\left\vert l\right\rangle , \\
\left\langle k\right\vert &=&\frac{1}{\sqrt{2\pi }}\sum\limits_{l=-\infty
}^{\infty }e^{-ikl}\left\langle l\right\vert ,
\end{eqnarray*}%
thus $\left\langle k_{1}\right\vert H_{S}\left\vert k_{2}\right\rangle =%
\mathcal{E}_{k_{1}}(b,\mathcal{A})\ \delta \left( k_{1}-k_{2}\right) $,
where 
\begin{equation}
\mathcal{E}_{k}(b,\mathcal{A})=\Omega \left\{ 1-\frac{\mathcal{A-}1}{%
\mathcal{A}}\sum_{n=0}^{\infty }\frac{\cos \left( b^{n}k\right) }{\mathcal{A}%
^{n}}\right\} \qquad ;\ -\pi <k<\pi .  \label{Ek}
\end{equation}

The eigenenergy $\mathcal{E}_{k}(b,\mathcal{A})$ can be related to lacunary
Taylor and Fourier series \cite{HMSch}. From (\ref{Ek}) we can define $%
\lambda (k)\equiv 1-\frac{1}{\Omega }\mathcal{E}_{k}(b,\mathcal{A})$, this
function obeys the scaling equation 
\begin{equation*}
\lambda (k)=\frac{1}{\mathcal{A}}\lambda (bk)+\frac{\mathcal{A-}1}{\mathcal{A%
}}\cos (k).
\end{equation*}%
It has been shown \cite{HMSch81} that the nonanalytic part $\lambda _{\text{s%
}}(k)$ of $\lambda (k)$ satisfies the homogeneous equation 
\begin{equation}
\lambda _{\text{s}}(k)=\frac{1}{\mathcal{A}}\lambda _{\text{s}}(bk),
\label{scaling}
\end{equation}%
so that 
\begin{equation*}
\lambda _{\text{s}}(k)=\left\vert k\right\vert ^{\mu }Q(k),\text{ with}\ \mu
=\ln \mathcal{A}/\ln b,
\end{equation*}%
here $Q(k)=Q(bk)$ is a bounded function periodic in $\ln \left\vert
k\right\vert $ with period $\ln b$ of quite intrincated structure \cite%
{HSM81}. Therefore our eigenenergy $\mathcal{E}_{k}(b,\mathcal{A})$ shares
some similarities with critical phenomena analysis (renormalization group
transformation). In addition if the parameter $b$ is an integer it has been
shown \cite{HMSch} that the power series in $z\equiv e^{ik}$ of $\lambda (k)$%
, has gaps or missing terms. These gaps lead, by using the Fabry's theorem,
to the concept of noncontinuability of the series of $\lambda (k)$, and so
to the conclusion that $\lambda (k)$ has extremely complicated behavior as a
function of $k$. In fact, for $\mu <1$, $\lambda (k)$ is Weierstrass'
example of a function which at no point possesses a finite derivative. When
this result is translated to the eigenenergy $\mathcal{E}_{k}(b,\mathcal{A})$
we may conclude that the density of states (DOS) is not well defined if $\mu
<1$ this will be shown also in the next subsection, see Eq. (\ref{diverge}).

As we mentioned before the type of Hamiltonians (\ref{H}), with shift
operators as presented in (\ref{a-general}) and (\ref{a-general2}), share
the properties of been diagonalized in Fourier space. The important point of
our Weierstrass quantum model is that the eigenenergy turns to be non
differentiable for the critical value $\mu <1$ (i.e., $b>A$), this fact will
be analyzed in the context of the time evolution of the reduced density
matrix, in detail, in the next sections.

An analysis concerning quantum walks with long-range steps (but without
dissipation) was carried out by M\"{u}lken et al. \cite{Blumen08a}, in that
paper it was shown that there exist a universal behavior for the quantum
walks which is different from the universality of long-range classical
random walks. In our present work we go one step forward and study a
long-range quantum walk coupled to a thermal bath, then new universalities
are found for the decoherence of the system.

\subsubsection{Density of states}

Using the Green function of the Hamiltonian $H_{S}$ the DOS can
straightforwardly be calculated if we known the matrix elements of $\mathbf{G%
}(Z)=\left[ Z-H_{S}\right] ^{-1}$, for example by using the formula $D(%
\mathcal{E})=\lim_{\delta \rightarrow 0^{+}}\frac{-1}{\pi }\mbox{Im}Tr\left[ 
\mathbf{G}(\mathcal{E}+i\delta )\right] .$ Nevertheless for the particular
case of our QLW Hamiltonian, and due to the fact that we already have an
analytic expression for the continuous energy eigenvalue $\mathcal{E}_{k}(b,%
\mathcal{A})$, it is more convenient here to calculate the DOS by the
alternative formula%
\begin{eqnarray}
D(\mathcal{E}) &=&\frac{1}{2\pi }\int_{-\pi }^{+\pi }\delta \left( \mathcal{E%
}-\mathcal{E}_{k}(b,\mathcal{A})\right) dk  \label{DOS} \\
&=&\frac{1}{2\pi }\sum_{k_{j}}\left[ \left. \frac{d\mathcal{E}_{k}(b,%
\mathcal{A)}}{dk}\right\vert _{k=k_{j}}\right] ^{-1},  \notag
\end{eqnarray}%
where $k_{j}=k_{j}(\mathcal{E})$ are the solutions of the transcendental
equation: $\mathcal{E}=\mathcal{E}_{k_{j}}(b,\mathcal{A})$. From (\ref{Ek})
we see that%
\begin{equation}
\frac{d\mathcal{E}_{k}(b,\mathcal{A})}{dk}=\Omega \frac{\mathcal{A-}1}{%
\mathcal{A}}\sum_{n=0}^{\infty }\left( \frac{b}{\mathcal{A}}\right) ^{n}\sin
\left( b^{n}k\right) ,  \label{diverge}
\end{equation}%
then we expect that when $b>\mathcal{A}$ the right-hand-side may diverge for
some values of $k_{j}$. This fact\ ultimately leads the DOS to be not well
defined, this issue can be seen as the occurrence of a complex null-set
inside the Brillouin zone, and is in agreement with the previous
mathematical report, coming from (\ref{scaling}), on the non
differentiability of $\lambda (k)$ when $\mu <1$. Only when $b<\mathcal{A}$
\ the eigenenergy $\mathcal{E}_{k}(b,\mathcal{A})$ is differentiable
anywhere and the behavior of the DOS for the Weierstrass' model shares
analogies with the DOS numerically calculated by M\"{u}lken et al. in a
long-range model \cite{Blumen08a}.

The Hamiltonian (\ref{H}) in the limit $b\rightarrow 1^{+\text{ }}$ is the
QRW model \cite{vK95,MOC-CH97,Blumen07}, then as expected, we can reobtain
from (\ref{Ek}) the usual \textit{tight-binding} eigenvalues $\mathcal{E}%
_{k}(b=1,\mathcal{A})=\Omega \left\{ 1-\cos \left( k\right) \right\} $, and
from (\ref{DOS}) the corresponding DOS:%
\begin{equation*}
D(\mathcal{E})=\frac{1}{\pi }\left( 2\Omega \mathcal{E}-\mathcal{E}%
^{2}\right) ^{-1/2},\qquad 0\leq \mathcal{E}\leq 2\Omega \qquad if\qquad b=1.
\end{equation*}

In conclusion: for $b>1$ the DOS follows from (\ref{DOS}), but this density
only is \textit{well defined} under the constraint $b/\mathcal{A}<1.$ In the
opposite case $b/\mathcal{A}>1$ the DOS is not defined because the function $%
\mathcal{E}_{k}(b,\mathcal{A})$ is non-differentiable. In figure 1 we have
plotted $\mathcal{E}_{k}(b,\mathcal{A})$\ for several values of Weierstrass'
parameters $b,\mathcal{A}$, in this plot the non differentiable signature of
the eigenenergy of the Levy walk Hamiltonian can clearly be seen.
\begin{figure}[t!]
\includegraphics[width=0.95 \columnwidth,clip]{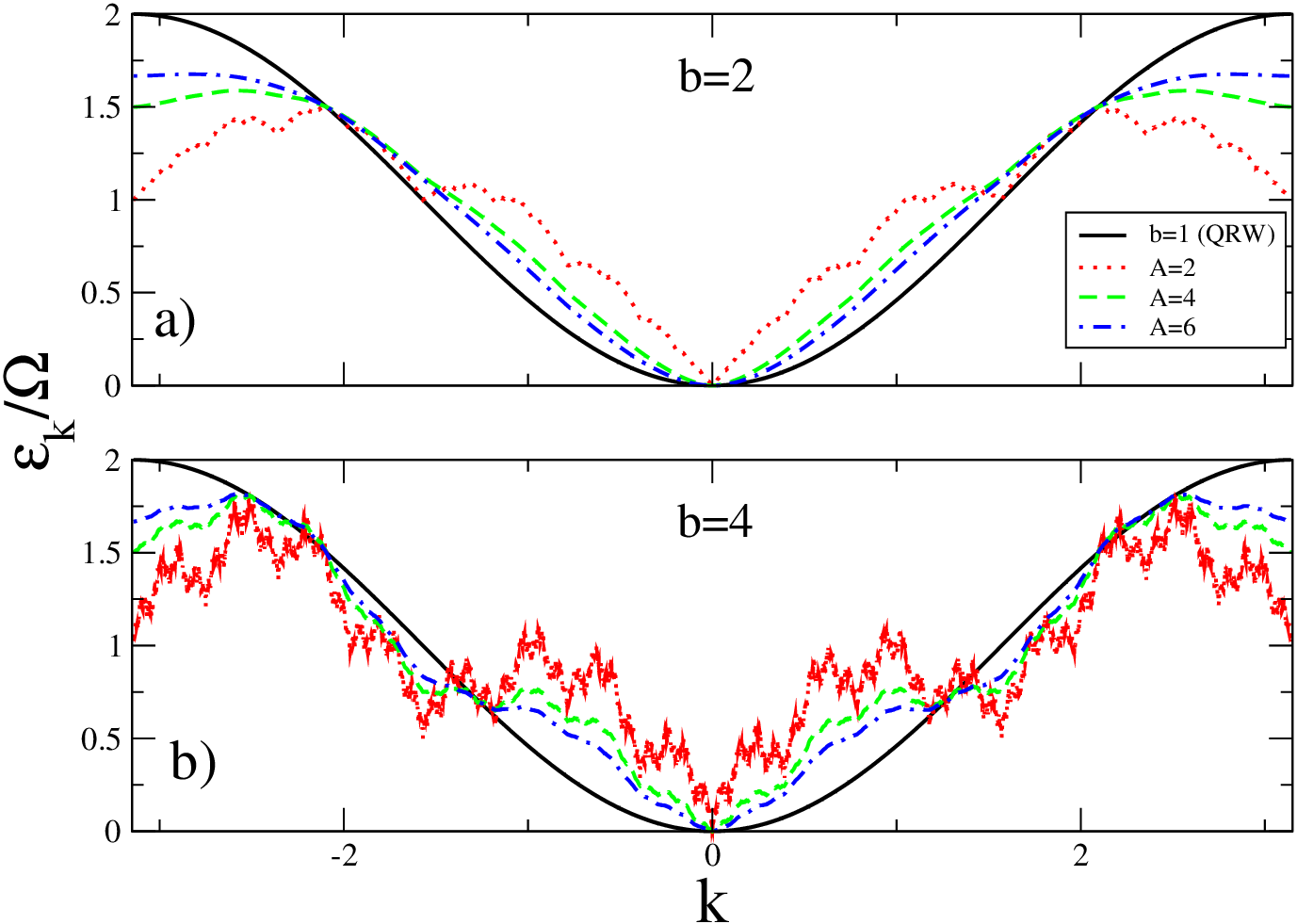}
\caption{Plot of $\mathcal{E}_{k}(b,\mathcal{A)}/\Omega $, the continuous
energy eigenvalue of Hamiltonian (\ref{H}) as a function of the Fourier
number $k$ in the first Brillouin zone (reciprocal to the Wannier lattice
index). In (a) and (b) the straight lines correspond to the \textit{%
tight-binding} case, i.e., the QRW model ($b=1$). The other lines (dotted,
slashed and dotted-slashed) correspond to the QLW for different values of
Weierstrass' parameters: $b(=2,4),\mathcal{A}(=2,4,6)$. For large values of
the Weierstrass' rate $b/\mathcal{A}>1$ the non-differentiable structure of $%
\mathcal{E}_{k}(b,\mathcal{A)}$ is clearly visible as was predicted from the
scaling (\ref{scaling}) and also from Eq. (\ref{diverge}).}
\label{fig1}
\end{figure}

Here it is important to note the difference with the \textit{classical}
characteristic function of the Weierstrass random walk (discrete space Levy
flight \cite{Mandelbrot,HSM81,MSh94,slesinger,libro}). In the classical walk
the condition to have an infinite mean square displacement per step is $%
b^{2}/\mathcal{A}>1$ (because the second moment is given by the second
derivative of $\lambda (k)$). Then when clustering occurs, the number of
subclusters within a distance $b^{n}$ of the origin is, on average, $%
\mathcal{A}^{n}$. Note that for the \textit{classical walk}, the
characteristic function of the Weierstrass walk is non-differentiable when $%
0<\mu <1$. Then it is possible to call\ to the value $\mu =\ln \mathcal{A}%
/\ln b$ the \textit{Hausdorff-Besicovitch dimension} of the set of sites
visited by a classical walk \cite{slesinger,HMSch}. In quantum mechanics a
similar analysis could be done in the context of the Sch\"{o}%
dringer-Langevin picture \cite{jpa}, i.e., without using the density matrix.

We want to remark that if $\mu <1$ ($b>A$) the eigenenergy $\mathcal{E}%
_{k}(b,\mathcal{A})$\ is non-differentiable and the DOS\ is not well
defined. A rigorous calculus based on the scaling properties (\ref{scaling})
of the eigenenergy leads to the conclusion that the record of $\mathcal{E}%
_{k}(b,\mathcal{A})$ as a function of $k$, for $b$ above the critical value $%
b>A$, is a self-affine function. In fact, a fractal dimension can be
measured (for example) by using the Box-counting technique and the
prediction gives $D=2-\mu $ for $0<\mu <1$ \cite{Mandelbrot2,Feder,libro}.
An alternative technique based in the analysis properties of the zero
crossing \cite{HIJ-2007} has recently been reported to be suitable to
characterizes processes that are continuous, but the \textit{derivative} has
fractal properties. Interestingly many years ago these processes were termed
\textquotedblleft subfractals\textquotedblright\ \cite{J1981}.

Note that we are assuming \textit{integer} values for $b$ to avoid
non-commensurability problems for the walk in the associated discrete
infinite dimensional Hilbert space $\mathbf{H}$ of lattice parameter $%
\epsilon =1$ (at the end of appendix B we present some discussion on the
case when the lattice parameter $\epsilon $ goes to the continuous limit).

\subsection{The Quantum Master Equation\ for the quantum Levy walk}

Concerning the quantum bath (see appendix A for details), we will assume
that the thermal bath $\mathcal{B}$ is an infinite set of oscillators, and
that the interaction with the phonon bath causes the free particle to jump
either to the right or to the left. Thus we consider the operators $V_{\beta
}$ appearing in the interaction Hamiltonian (\ref{hi}) to be proportional to
the Weierstrass shift operators $a$ and $a^{\dag }$ defined in Eq. (\ref%
{Shift}). In the present paper we will study a particular coupling with the
heat bath, nevertheless going back to appendices A, B and C it is
straightforward to write down the QME\ considering any other type of
coupling. For example: if $V_{\beta }\propto a-a^{\dag }$ we can model an
interaction that in the continuos limit (lattice parameter going to zero)
coupled the velocity of the free particle with the heat bath.

Taking into account that $[a^{\dag },a]=0$ we trivially get that in the
Heisenberg representation Weierstrass's shift operator does not have a time
evolution, i.e.,:%
\begin{equation*}
V_{\beta }(-\tau )\equiv \exp (-i\tau H_{S}/\hbar )V_{\beta }\exp (i\tau
H_{S}/\hbar )=V_{\beta }(0).
\end{equation*}%
This result ultimately will lead to the fact that the Kossakowski-Lindblad
(KL) generator \cite{kossa,lindblad} will be completely positive \cite{jpa}.
Thus, using Eqs. (\ref{eq3c}) and (\ref{eq4b}) in (\ref{eq2b}), we can write
the Quantum Master Equation (QME), for the reduced density matrix $\rho $,
in the form (see (\ref{QMEfull}) in appendix C)

\begin{equation}
\dot{\rho}=\frac{-i}{\hbar }\left[ H_{eff},\rho \right] +\frac{\pi \alpha }{%
4\beta \hbar }\left( 2a\rho a^{\dag }-a^{\dag }a\rho -\rho a^{\dag }a\right)
+\frac{\pi \alpha }{4\beta \hbar }\left( 2a^{\dag }\rho a-aa^{\dag }\rho
-\rho aa^{\dag }\right) .  \label{QMEa}
\end{equation}%
Here $\alpha >0$ is the dissipative constant and $\beta $ the inverse of the
temperature of the bath.

Note that in the case $b\neq 1$ we have $a^{\dag }a\neq \mathbf{1}$ (see
appendix B) therefore the effective Hamiltonian has a non-trivial
contribution:%
\begin{equation}
H_{eff}=H_{S}-\hbar \omega _{c}a^{\dag }a,\quad \omega _{c}>0.  \label{Heff}
\end{equation}%
The effective Hamiltonian can also be diagonalized in the Fourier basis,
therefore we get%
\begin{eqnarray}
\left\langle k_{1}\right\vert H_{eff}\left\vert k_{2}\right\rangle &=&\left[ 
\mathcal{E}_{k_{1}}(b,\mathcal{A)}\ -\hbar \omega _{c}\left( \frac{\mathcal{%
A-}1}{\mathcal{A}}\right) ^{2}\sum\limits_{n_{2},n_{1}=0}^{\infty }\frac{%
\cos k_{1}\left( b^{n_{2}}-b^{n_{1}}\right) }{\mathcal{A}^{n_{1}+n_{2}}}%
\right] \delta \left( k_{1}-k_{2}\right)  \notag \\
&\equiv &\mathrm{E}_{k_{1}}(b,\mathcal{A)}\ \delta \left( k_{1}-k_{2}\right)
,  \label{He1}
\end{eqnarray}%
where we have used (\ref{aaK}) and $\mathcal{E}_{k_{1}}(b,\mathcal{A)}$ is
given in (\ref{Ek}). Note that $\hbar \omega _{c}$ is an upper bound energy
which is characteristic of the Ohmic approximation \cite{caldeira}, (see (%
\ref{Hef}) in the appendix C). In figure 2 we have plotted $\mathrm{E}_{k}(b,%
\mathcal{A)}$\ for several values of Weierstrass' parameters $b,\mathcal{A}$%
, in this plot it can be seen that the effective eigenenergy associated to
the Hamiltonian (\ref{Heff}) has larger fluctuations than $\mathcal{E}%
_{k_{1}}(b,\mathcal{A)}$ the eigenenergy associated to the naked Levy walk
Hamiltonian.
\begin{figure}[t!]
\includegraphics[width=0.95 \columnwidth,clip]{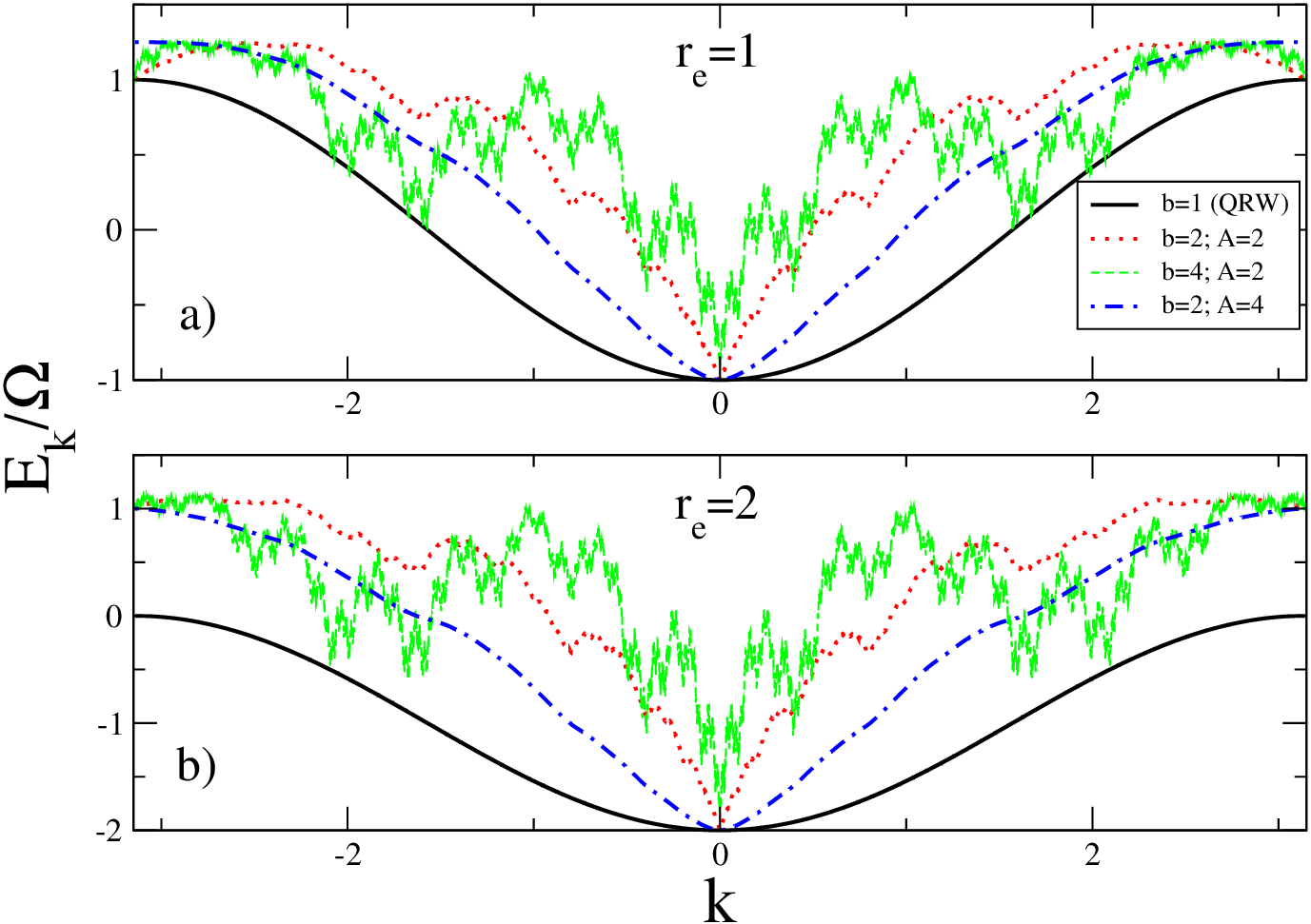}
\caption{Plot of $\mathrm{E}_{k_{1}}(b,\mathcal{A)}/\Omega $, the continuous
energy eigenvalue of the effective Hamiltonian (\ref{Heff}) as a function of
the Fourier number $k$ in the first Brillouin zone. In (a) the energy rate
is $r_{e}=\hbar \omega _{c}/\Omega =1$, and in (b) is $r_{e}=2$. The
straight lines correspond to the \textit{tight-binding} case ($b=1$). Other
lines (dotted, slashed and dotted-slashed) correspond to the QLW\ model for
different values of Weierstrass' parameters $b(=2,4),\mathcal{A}(=2,4)$. The
physical meaning of the contribution $\hbar \omega _{c}a^{\dag }a$ in the
effective Hamiltonian $H_{eff}$ can be thought as increasing the
multiplicity of the hopping structure, see (\ref{aal1l2}). For $b>A$ and
large values of the rate $r_{e}$ the amplitude of the non-differential
structure get larger.}
\label{fig2}
\end{figure}

When $b=1$ (the QRW case) the effective Hamiltonian only introduces a
trivial constant because $\lim_{b\rightarrow 1}a^{\dag }a\rightarrow \mathbf{%
1}$. Therefore, for $b=1$ the KL generator has a very simple expression,
form (\ref{QMEa}) the QME results \cite{vK95,MOC-CH97}%
\begin{equation*}
L\left[ \rho \right] =\frac{-i}{\hbar }\left[ H_{S},\rho \right] +\frac{\pi
\alpha }{2\beta \hbar }\left( a\rho a^{\dag }+a^{\dag }\rho a\right) -\frac{%
\pi \alpha }{\beta \hbar }\rho \qquad \text{if\qquad }b=1,
\end{equation*}%
the group $\pi \alpha /2\beta \hbar $ can be called the diffusion constant $%
D $ and is given in units of [time]$^{-1}$ because the lattice parameter is
dimensionless. The case $D\rightarrow \infty $ corresponds to the high
temperature limit, see appendix C.

\subsection{\protect\bigskip On the second moment of the quantum Levy walk}

From the QME (\ref{QMEa}) we can obtain the dynamics of any operator, in
particular here we are interested in the evolution of the dispersion of the
position operator $\mathbf{q}$, which in the Wannier basis has the matrix
elements:%
\begin{equation}
\left\langle l_{1}\right\vert \mathbf{q}\left\vert l_{2}\right\rangle
=l\delta _{l_{1},l_{2}},  \label{position}
\end{equation}%
note that $\mathbf{q}$ is defined as a dimensionless position operator. In
the $\left\vert l\right\rangle $ representation it is possible to see that
the thermal mean-value time-evolution of the first and second quantum
moments can be written in the form: 
\begin{eqnarray}
\frac{d}{dt}\left\langle \mathbf{q}(t)\right\rangle &=&\frac{d}{dt}Tr\left\{ 
\mathbf{q}(t)\rho (0)\right\} =Tr\left\{ \mathbf{q}\dot{\rho}(t)\right\}
\label{q1} \\
&=&\frac{-i}{\hbar }Tr\left\{ \rho \left[ \mathbf{q},H_{eff}\right] \right\}
\notag \\
\frac{d}{dt}\left\langle \mathbf{q}^{2}(t)\right\rangle &=&Tr\left\{ \mathbf{%
q}^{2}\dot{\rho}(t)\right\}  \label{q2} \\
&=&\frac{-i}{\hbar }Tr\left\{ \rho \left[ \mathbf{q}^{2},H_{eff}\right]
\right\} +\frac{\pi \alpha }{\beta \hbar }\left( \frac{\mathcal{A-}1}{%
\mathcal{A}}\right) ^{2}\sum_{n_{1},n_{2}=0}^{\infty }\sum_{l=-\infty
}^{\infty }\frac{b^{n_{1}}b^{n_{2}}}{\mathcal{A}^{n_{1}+n_{2}}}\left\langle
l\right\vert \rho (t)\left\vert l-b^{n_{1}}+b^{n_{2}}\right\rangle .  \notag
\end{eqnarray}

Setting $b=1$ in (\ref{q1}) and (\ref{q2}), i.e., in the limit of the usual
QRW, we can write in the Heisenberg representation:%
\begin{eqnarray}
\frac{d}{dt}\mathbf{q}(t) &=&\frac{-i}{\hbar }\left[ \mathbf{q},H_{S}\right]
\qquad if\qquad b=1,  \label{1mom} \\
\frac{d}{dt}\mathbf{q}^{2}(t) &=&\frac{-i}{\hbar }\left[ \mathbf{q}^{2},H_{S}%
\right] +\frac{\pi \alpha }{\beta \hbar }\qquad if\qquad b=1.  \label{2mom}
\end{eqnarray}%
These equations can easily be solved. Using that $\left[ \mathbf{q},H_{S}%
\right] =\frac{\Omega }{2}\left( a-a^{\dag }\right) $ (see appendix B), and
that $a(t)=a(0),a^{\dag }(t)=a^{\dag }(0)$ we get for the time evolution of
the position operator:%
\begin{equation*}
\mathbf{q}(t)=\frac{-i\Omega }{2\hbar }\left( a-a^{\dag }\right) t+\mathbf{q}%
(0)\qquad if\qquad b=1.
\end{equation*}%
In the same way, the variance of the QRW (the NN walk model) can be
calculated giving: $\left\langle \mathbf{q}(t)^{2}\right\rangle
-\left\langle \mathbf{q}(t)\right\rangle ^{2}=\frac{1}{2}\left( \frac{\Omega
t}{\hbar }\right) ^{2}+2Dt$, with $2D\equiv $ $\frac{\pi \alpha }{\beta
\hbar }$, which is the expected dissipative result \cite%
{MOC-CH97,kampen,libro,vK95}. From (\ref{1mom}) and (\ref{2mom}) it is
possible to see that von Neumann's term gives a contribution of the form $%
\propto t^{2}$ for the time-evolution of the second moment, this is a well
known quantum result \cite{Blumen08a}, see eq. (\ref{q2det}) for a general
discussion. The dissipative contribution comes from the interaction with the
bath $\mathcal{B}$, giving the classical diffusive behavior $\propto t$.

From (\ref{q2}) and due to the coherent dynamics involved through the
time-evolution of the off-diagonal elements $\left\langle l\right\vert \rho
(t)\left\vert l-b^{n_{1}}+b^{n_{2}}\right\rangle $, it is not simple to
realize what will be the dynamics in the general case when $b\neq 1$. But in
principle any higher moments of $\mathbf{q}(t)$ can also be analyzed in the
same way from our QME (\ref{QMEa}).

Let us now analyze the case $b\neq 1$. In this case the interaction with $%
\mathcal{B}$ produces long-range hopping, and consistently a non-trivial
quantum decoherence phenomenon. The dissipative term of (\ref{q2}), can also
be written in the form:%
\begin{equation}
\frac{\pi \alpha }{\beta \hbar }\left( \frac{\mathcal{A-}1}{\mathcal{A}}%
\right) ^{2}\sum_{n_{1},n_{2}=0}^{\infty }\left( \frac{b}{\mathcal{A}}%
\right) ^{n_{1}+n_{2}}\sum_{l=-\infty }^{\infty }\left\langle
l+b^{n_{1}}\right\vert \rho (t)\left\vert l+b^{n_{2}}\right\rangle .
\label{non-D}
\end{equation}%
We can explicitly calculate this contribution by going to the Fourier
representation. First of all, here we will assume that the initial condition
for the reduced density matrix was prepared in a pure Wannier state: $\rho
(0)=\left\vert l_{0}\right\rangle \left\langle l_{0}\right\vert $, so%
\begin{equation}
\rho (0)_{k_{1},k_{2}}\equiv \left\langle k_{1}\right\vert \rho
(0)\left\vert k_{2}\right\rangle =\frac{1}{2\pi }\exp \left(
-i(k_{1}-k_{2})l_{0}\right) ,  \label{pure}
\end{equation}%
thus $\left\langle k\right\vert \rho (0)\left\vert k\right\rangle =\frac{1}{%
2\pi }$. On the other hand, from (\ref{QMEa}) is possible to see that $\frac{%
d}{dt}\left\langle k\right\vert \rho (t)\left\vert k\right\rangle =0$
therefore $\left\langle k\right\vert \rho (t)\left\vert k\right\rangle =%
\frac{1}{2\pi },\forall t$, (see next section). Now going back to (\ref%
{non-D}) we can write:%
\begin{eqnarray*}
(\ref{non-D}) &=&\frac{\pi \alpha }{\beta \hbar }\left( \frac{\mathcal{A-}1}{%
\mathcal{A}}\right) ^{2}\int \int dk_{1}dk_{2}\sum_{n_{1},n_{2}=0}^{\infty
}\left( \frac{b}{\mathcal{A}}\right) ^{^{n_{1}+n_{2}}}\sum_{l=-\infty
}^{\infty }\left\langle l+b^{n_{1}}\right\vert \left. k_{1}\right\rangle
\left\langle k_{1}\right\vert \rho (t)\left\vert k_{2}\right\rangle
\left\langle k_{2}\right. \left\vert l+b^{n_{2}}\right\rangle \\
&=&\frac{\pi \alpha }{\beta \hbar }\left( \frac{\mathcal{A-}1}{\mathcal{A}}%
\right) ^{2}\int \int dk_{1}dk_{2}\sum_{n_{1},n_{2}=0}^{\infty }\left( \frac{%
b}{\mathcal{A}}\right)
^{^{n_{1}+n_{2}}}e^{ib^{n_{1}}k_{1}-ib^{n_{2}}k_{2}}\left\langle
k_{1}\right\vert \rho (t)\left\vert k_{2}\right\rangle \sum_{l=-\infty
}^{\infty }\frac{e^{i(k_{1}-k_{2})l}}{2\pi } \\
&=&\frac{\pi \alpha }{\beta \hbar }\left( \frac{\mathcal{A-}1}{\mathcal{A}}%
\right) ^{2}\sum_{n_{1},n_{2}=0}^{\infty }\left( \frac{b}{\mathcal{A}}%
\right) ^{^{n_{1}+n_{2}}}\int_{-\pi }^{\pi
}dk_{1}e^{i(b^{n_{1}}-b^{n_{2}})k_{1}}\left\langle k_{1}\right\vert \rho
(t)\left\vert k_{1}\right\rangle \\
&=&\frac{\pi \alpha }{\beta \hbar }\left( \frac{\mathcal{A-}1}{\mathcal{A}}%
\right) ^{2}\sum_{n_{1},n_{2}=0}^{\infty }\left( \frac{b}{\mathcal{A}}%
\right) ^{^{n_{1}+n_{2}}}\frac{1}{2\pi }\int_{-\pi }^{\pi
}dk_{1}e^{i(b^{n_{1}}-b^{n_{2}})k_{1}} \\
&=&\frac{\pi \alpha }{\beta \hbar }\left( \frac{\mathcal{A-}1}{\mathcal{A}}%
\right) ^{2}\sum_{n_{1},n_{2}=0}^{\infty }\left( \frac{b}{\mathcal{A}}%
\right) ^{^{n_{1}+n_{2}}}\frac{\sin (b^{n_{1}}-b^{n_{2}})\pi }{%
(b^{n_{1}}-b^{n_{2}})\pi },
\end{eqnarray*}%
noting that we have assumed $b=\mbox{integer}$ we can write:%
\begin{eqnarray}
(\ref{non-D}) &=&\frac{\pi \alpha }{\beta \hbar }\left( \frac{\mathcal{A-}1}{%
\mathcal{A}}\right) ^{2}\left( \frac{\mathcal{A}^{2}}{\mathcal{A}^{2}-b^{2}}%
+\sum_{n_{1}\neq n_{2}}^{\infty }\left( \frac{b}{\mathcal{A}}\right)
^{^{n_{1}+n_{2}}}\frac{\sin (b^{n_{1}}-b^{n_{2}})\pi }{(b^{n_{1}}-b^{n_{2}})%
\pi }\right)  \notag \\
&=&\left\{ 
\begin{array}{c}
\!\frac{\pi \alpha }{\beta \hbar },\quad if\quad b=1, \\ 
\qquad \frac{\pi \alpha }{\beta \hbar }\frac{\left( \mathcal{A-}1\right) ^{2}%
}{\mathcal{A}^{2}-b^{2}},\quad if\quad \mathcal{A}>1,b(\text{integer})<%
\mathcal{A}, \\ 
\!\infty ,\quad if\quad b>\mathcal{A}.\qquad \qquad \blacksquare%
\end{array}%
\right.  \label{dqdt2}
\end{eqnarray}%
The expression (\ref{dqdt2}) shows that it is only in the case when $b>$ $%
\mathcal{A}$ that a divergent behavior for the thermal second moment of the
QLW, $\left\langle \mathbf{q}^{2}(t)\right\rangle ,$ can arise. This quantum
result is quite different from the classical clustering Levy flight
counterpart ($b^{2}>\mathcal{A}$) \cite%
{Mandelbrot,HSM81,MSh94,slesinger,libro}.

The explicit solution of $\left\langle \mathbf{q}^{2}(t)\right\rangle $ can
alternatively be obtained calculating $Tr\left[ \mathbf{q}^{2}\rho (t)\right]
$ (this is done in appendix D):

\begin{equation}
\left\langle \mathbf{q}^{2}(t)\right\rangle =\left\{ 
\begin{array}{c}
\frac{1}{2}\left( \frac{\Omega }{\hbar }t\right) ^{2}+\frac{\pi \alpha }{%
\beta \hbar }t+l_{0}^{2},\quad if\quad \mathcal{A}>1,b=1, \\ 
\\ 
\frac{\left( \mathcal{A-}1\right) ^{2}}{\mathcal{A}^{2}-b^{2}}\left( \frac{1%
}{2}\frac{\Omega ^{2}}{\hbar ^{2}}t^{2}+\frac{2\omega _{c}^{2}\left(
b-1\right) ^{2}}{\left( 1-1/\mathcal{A}^{2}\right) \left( 1-b/\mathcal{A}%
^{2}\right) ^{2}}t^{2}+\frac{\pi \alpha }{\beta \hbar }t\right)
+l_{0}^{2},\quad if\quad \mathcal{A}>1,b<\mathcal{A}.%
\end{array}%
\right.  \label{q2det}
\end{equation}%
By taking the limit of null dissipation, i.e., $\alpha =0$ we reobtain the
quantum behavior $\left\langle \mathbf{q}^{2}(t)\right\rangle \propto t^{2}$
for $b<\mathcal{A}$, this result is in agreement with previous reports on
the universal behavior of quantum walks with long-range steps \cite%
{Blumen08a}.

We conclude this section noting that the quantum coherence enlarges the
threshold (compared to the classical one) to have a finite second moment for
the walk. For example, if $b=2,\mathcal{A}=3$ the classical second moment is
not defined, but\ the thermal quantum average $\left\langle \mathbf{q}%
^{2}(t)\right\rangle $ is finite!

\subsection{Time evolution of the density matrix}

\subsubsection{\protect\bigskip Off-diagonal relaxation}

The QME (\ref{QMEa}) can be solved in the basis of the eigenvectors of $%
H_{S} $. Introducing the representation $\left\vert k\right\rangle $, and
using that $\left\langle k_{1}\right\vert H_{eff}\left\vert
k_{2}\right\rangle =\mathrm{E}_{k_{1}}\ \delta \left( k_{1}-k_{2}\right) $,
with \textrm{E}$_{k_{1}}\equiv \mathrm{E}_{k_{1}}(b,\mathcal{A})$ given in (%
\ref{He1}) we get (see appendix C):%
\begin{eqnarray}
\left\langle k_{1}\right\vert \dot{\rho}\left\vert k_{2}\right\rangle &=&%
\frac{-i}{\hbar }\left( \mathrm{E}_{k_{1}}-\mathrm{E}_{k_{2}}\right)
\left\langle k_{1}\right\vert \rho \left\vert k_{2}\right\rangle  \notag \\
&&+\frac{\pi \alpha }{2\beta \hbar }\left( \frac{\mathcal{A-}1}{\mathcal{A}}%
\right) ^{2}\sum_{n_{1},n_{2}=0}^{\infty }\frac{1}{\mathcal{A}^{n_{1}+n_{2}}}%
\left[ 2\cos \left( k_{1}b^{n_{1}}-k_{2}b^{n_{2}}\right) \right.  \notag \\
&&-\left. \cos k_{1}\left( b^{n_{1}}-b^{n_{2}}\right) -\cos k_{2}\left(
b^{n_{1}}-b^{n_{2}}\right) \right] \left\langle k_{1}\right\vert \rho
\left\vert k_{2}\right\rangle .  \label{evo1}
\end{eqnarray}

From (\ref{evo1}) the general solution for the reduced density matrix is%
\begin{equation}
\left\langle k_{1}\right\vert \rho (t)\left\vert k_{2}\right\rangle =\rho
(0)_{k_{1}k_{2}}\ \exp \left( \mathcal{F}\left( k_{1},k_{2},b,\mathcal{A}%
\right) t\right) ,  \label{evo2}
\end{equation}%
with 
\begin{eqnarray}
\mathcal{F}\left( k_{1},k_{2},b,\mathcal{A}\right) &\equiv &\frac{i}{\hbar }%
\left[ \Omega \frac{\mathcal{A-}1}{\mathcal{A}}\sum_{n=0}^{\infty }\frac{1}{%
\mathcal{A}^{n}}\left( \cos \left( b^{n}k_{1}\right) -\cos \left(
b^{n}k_{2}\right) \right) \right.  \notag \\
&&+\left. \hbar \omega _{c}\left( \frac{\mathcal{A-}1}{\mathcal{A}}\right)
^{2}\sum\limits_{n_{2},n_{1}=0}^{\infty }\frac{\cos k_{1}\left(
b^{n_{2}}-b^{n_{1}}\right) }{\mathcal{A}^{n_{1}+n_{2}}}-\frac{\cos
k_{2}\left( b^{n_{2}}-b^{n_{1}}\right) }{\mathcal{A}^{n_{1}+n_{2}}}\right] 
\notag \\
&&+\frac{\pi \alpha }{2\beta \hbar }\left( \frac{\mathcal{A-}1}{\mathcal{A}}%
\right) ^{2}\sum_{n_{1},n_{2}=0}^{\infty }\frac{1}{\mathcal{A}^{n_{1}+n_{2}}}%
\left[ 2\cos \left( k_{1}b^{n_{1}}-k_{2}b^{n_{2}}\right) \right.  \notag \\
&&-\left. \cos k_{1}\left( b^{n_{1}}-b^{n_{2}}\right) -\cos k_{2}\left(
b^{n_{1}}-b^{n_{2}}\right) \right] ,  \label{evo3}
\end{eqnarray}%
note that $\mathcal{F}\left( k,k,b,\mathcal{A}\right) =0.$

For the QRW (case $b=1$) the QME (\ref{evo1}) reduces to a simpler form%
\begin{equation*}
\left\langle k_{1}\right\vert \dot{\rho}\left\vert k_{2}\right\rangle =\left[
\frac{-i}{\hbar }\left( \mathrm{E}_{k_{1}}-\mathrm{E}_{k_{2}}\right) +\frac{%
\pi \alpha }{\beta \hbar }\left( \cos \left( k_{1}-k_{2}\right) -1\right) %
\right] \left\langle k_{1}\right\vert \rho \left\vert k_{2}\right\rangle ,
\end{equation*}%
where $\mathrm{E}_{k}=\Omega \left\{ 1-\cos k\right\} -\hbar \omega _{c}.$
Note that in the NN case, the off-diagonal relaxation of the element $\rho
_{k,0}(t)$ (from the homogeneous Fourier mode) is just controlled by a
trivial trigonometric function: 
\begin{equation*}
\mathcal{F}\left( k,0,b=1,\mathcal{A}\right) =-\left[ i\frac{\Omega }{\hbar }%
+\frac{\pi \alpha }{\beta \hbar }\right] \left( 1-\cos k\right) .
\end{equation*}%
This result resembles the relaxation of the Fourier modes of a classical one
dimensional ordered walk with a diffusion coefficient $2D=\frac{\pi \alpha }{%
\beta \hbar }$ (see appendix E). Note however that in quantum mechanic the
relaxation of the initial position is controlled by a double Fourier
integral, so we cannot expect the same long-time asymptotic behavior as in
classic (see next sections for details).

\subsubsection{Diagonal relaxation}

\bigskip

In general for $b\geq 1,$ by putting $k_{1}=k_{2}$ in (\ref{evo1}), we see
that the diagonal elements $\rho _{kk}(t)$ remain constant in time:%
\begin{equation}
\frac{d}{dt}\left\langle k\right\vert \rho \left\vert k\right\rangle =0.
\label{pkk}
\end{equation}%
This result tell that even when there is a decoherence in the off-diagonal
elements of the density matrix, the probability distribution of the Fourier
modes is invariant in time. The physical interpretation of this fact can be
understood by noting that the thermal average of the kinetic energy is
constant in time. As we mention before this is a consequence of the general
form of the shift operators (\ref{a-general}) and (\ref{a-general2}).

To prove this fact for the Weierstrass model, we first define a
pseudo-momentum\ operator:%
\begin{equation}
\mathbf{p}\equiv \frac{m}{i\hbar }\left[ \mathbf{q},H_{S}\right] ,  \label{p}
\end{equation}%
where $m$ represents the mass of our free particle in the lattice. From (\ref%
{qH}) it is simple to see that $\mathbf{p}$ is diagonal in the Fourier basis 
$\mathbf{p}\left\vert k\right\rangle =\mathbf{p}_{k}\left\vert
k\right\rangle $, i.e.,

\begin{eqnarray}
\left\langle k_{1}\right\vert \mathbf{p}\left\vert k\right\rangle &=&\frac{%
m\Omega }{\hbar }\left( \frac{\mathcal{A-}1}{\mathcal{A}}\right)
\sum_{n}\left( \frac{b}{\mathcal{A}}\right) ^{n}\sin \left( kb^{n}\right)
\delta \left( k-k_{1}\right)  \label{momento} \\
&\equiv &\delta \left( k-k_{1}\right) \mathbf{p}_{k}.  \notag
\end{eqnarray}%
Once again the eigenvalue $\mathbf{p}_{k}$ is well defined only below the
threshold $b<\mathcal{A}$.

Assuming the system was prepared in the pure state: $\rho (0)=\left\vert
l_{0}\right\rangle \left\langle l_{0}\right\vert $, the thermal average of
the pseudo-momentum gives $\left\langle \mathbf{p}(t)\right\rangle
=0,\forall t$, and for $\left\langle \mathbf{p}^{2}(t)\right\rangle $ we get%
\begin{eqnarray}
\left\langle \mathbf{p}^{2}(t)\right\rangle &=&Tr\left[ \mathbf{p}^{2}\rho
(t)\right] =\int_{-\pi }^{\pi }dk\int_{-\pi }^{\pi }dk_{1}\left\langle
k\right\vert \mathbf{p}^{2}\left\vert k_{1}\right\rangle \left\langle
k_{1}\right\vert \rho (t)\left\vert k\right\rangle  \label{termalP} \\
&=&\int_{-\pi }^{\pi }dk\int_{-\pi }^{\pi }dk_{1}\ \left( \mathbf{p}%
_{k}\right) ^{2}\delta \left( k_{1}-k\right) \left\langle k_{1}\right\vert
\rho (t)\left\vert k\right\rangle  \notag \\
&=&\frac{1}{2\pi }\int_{-\pi }^{\pi }dk\ \left( \mathbf{p}_{k}\right) ^{2}=%
\text{constant,}  \notag
\end{eqnarray}%
where we have used that $\left\langle k\right\vert \rho (t)\left\vert
k\right\rangle =\frac{1}{2\pi },\forall t.$ In general it is possible to
prove that the thermal average of any observable which is diagonal in the
Fourier basis, will be constant in time.

It is interesting to comment here that for the QRW (the NN case) the
pseudo-momentum\ operator (\ref{p}) is in fact a discrete version of the
momentum, and can be written in the form (see appendix B) 
\begin{equation*}
\mathbf{p}\equiv \frac{\Omega m}{i\hbar 2}\left( a-a^{\dag }\right) ,\qquad 
\text{if\qquad }b=1,
\end{equation*}%
then it is simple to see that $\mathbf{\dot{p}}\equiv \frac{1}{i\hbar }\left[
\mathbf{p},H_{S}\right] =0$, so the \textit{discrete} momentum in the
Heisenberg representation is constant in time.\ The \textquotedblleft
lattice\textquotedblright\ commutation relation between the position and the
momentum gives%
\begin{equation*}
\left[ \mathbf{q},\mathbf{p}\right] \equiv -\frac{\Omega m}{i\hbar 2}\left(
a+a^{\dag }\right) ,\qquad \text{if\qquad }b=1.
\end{equation*}%
Thus in the continuous limit (taking the lattice parameter $\epsilon
\rightarrow 0$, see appendix B) we reobtain the usual commutation relation $%
\left[ \mathbf{q},\mathbf{p}\right] \equiv i\hbar \mathbf{1}$, \cite%
{MOC-CH97}.

\subsubsection{Quantum decoherence from a pure state}

In order to study the quantum decoherence due to the interaction with the
thermal bath $\mathcal{B}$, we propose here to analyze the time evolution of
the density matrix assuming that at time $t=0$ the system was prepared in a
pure state: $\rho (0)=\left\vert l_{0}\right\rangle $ $\left\langle
l_{0}\right\vert $. Then the quantum probability to be at site $l$ at time $%
t $ is given by%
\begin{eqnarray}
\left\langle l\right\vert \rho (t)\left\vert l\right\rangle &=&\int_{-\pi
}^{\pi }dk_{1}\int_{-\pi }^{\pi }dk_{2}\left\langle l\right. \left\vert
k_{1}\right\rangle \left\langle k_{1}\right\vert \rho (t)\left\vert
k_{2}\right\rangle \left\langle k_{2}\right\vert \left. l\right\rangle
\label{Pl} \\
&=&\frac{1}{2\pi }\int_{-\pi }^{\pi }dk_{1}\int_{-\pi }^{\pi }dk_{2}\exp %
\left[ i\left( k_{1}-k_{2}\right) l\right] \left\langle k_{1}\right\vert
\rho (t)\left\vert k_{2}\right\rangle  \notag \\
&=&\left( \frac{1}{2\pi }\right) ^{2}\int_{-\pi }^{\pi }dk_{1}\int_{-\pi
}^{\pi }dk_{2}\exp \left[ i\left( k_{1}-k_{2}\right) (l-l_{0})\right] \exp
\left( \mathcal{F}\left( k_{1},k_{2},b,\mathcal{A}\right) t\right) .  \notag
\end{eqnarray}%
Calling $(l-l_{0})=\Delta l$ the distance form the initial condition, we can
plot the probability $\mathit{P}(\Delta l,t)\equiv \left\langle l\right\vert
\rho (t)\left\vert l\right\rangle $ as a function of $\Delta l$ for
different values of time $t$, and Weierstrass' parameters $b,\mathcal{A}$.
Note that $\mathit{P}(\Delta l,t)\leq 1$ as expected because the density
matrix is well normalized $\sum_{l}\left\langle l\right\vert \rho
(t)\left\vert l\right\rangle =1,\forall t\neq \infty $, as can easily be
checked from (\ref{Pl}). In figure 3 we plot $\mathit{P}(\Delta l,t)$ for
different values of the frequency rates $r=\frac{\Omega }{\hbar }/D$, $%
r_{c}=\omega _{c}/D$ and Weierstrass' parameters $b,\mathcal{A}$, the non
diffusive characteristics of the profile $\mathit{P}(\Delta l,t)$ as well as
its trimodality behavior can clearly be seen. This trimodality is the result
of the combination between the non-diffusive characteristics (clustering) of
the Levy walk (by increasing $b>A$) and the quantum oscillations (coherence
by increasing $r>1,r_{c}>1$). All Fourier integrals were made using the
numerical integration quadrature method \cite{numerical}.
\begin{figure}[t!]
\includegraphics[width=0.95 \columnwidth,clip]{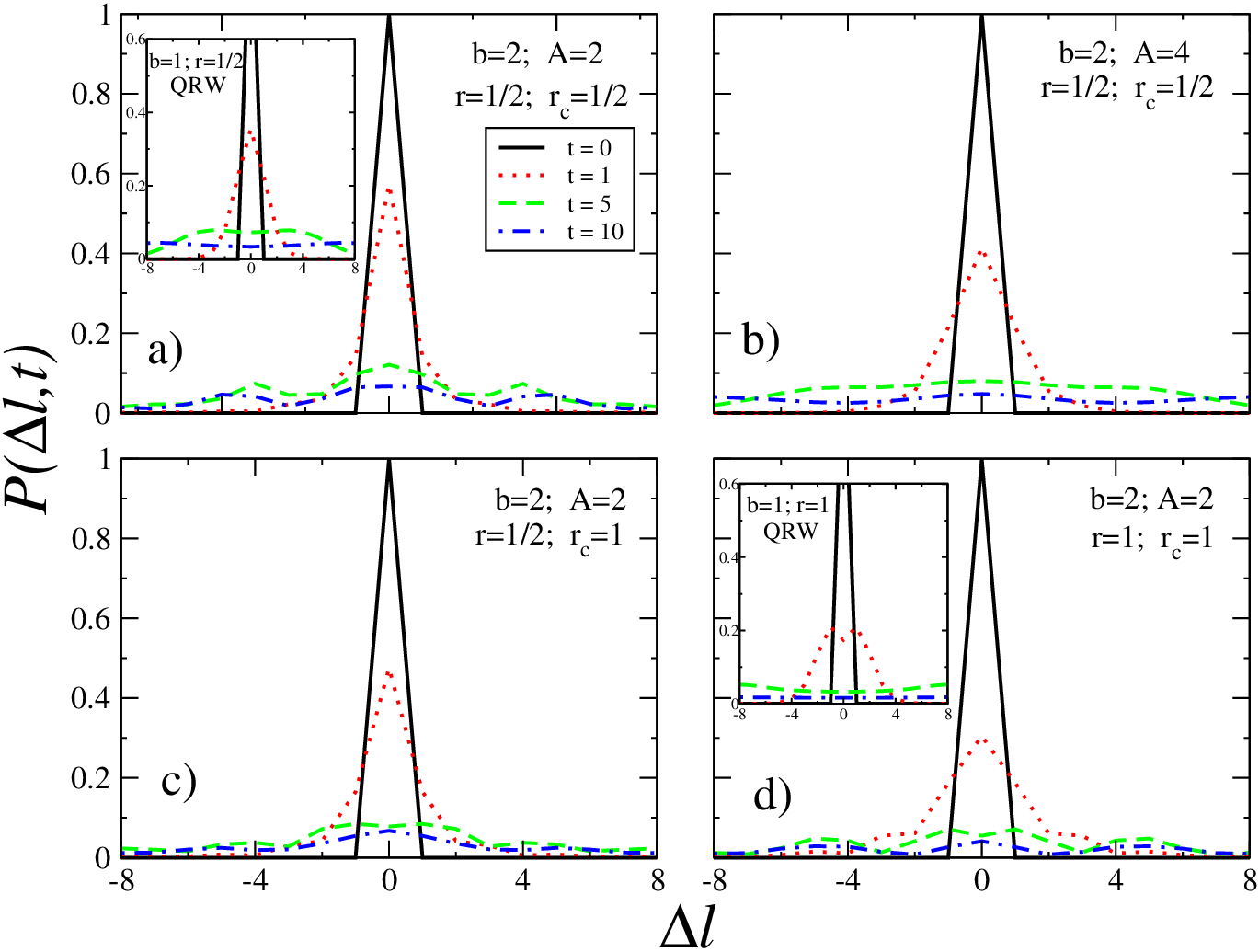}
\caption{Plot of the position probability $\mathit{P}(\Delta l,t)$ when the
system is prepared in the pure state $\rho (0)=\left\vert l_{0}\right\rangle
\left\langle l_{0}\right\vert $ as a function of $\Delta l$ (dimensionless
space difference $\Delta l\equiv l-l_{0}$), for several values of
dimensionless time $t$ ($D$ is given in units of [time]$^{-1}$). In the
insets we show the QRW case $(b=1)$, then it is possible to see that by
increasing the frequency rate $r=\frac{\Omega }{\hbar }/D$ the quantum
time-oscillations around $\Delta l=0$ persist longer before the profile
spreads in a diffusive way. In figures (a)-(d) we show the same profile but
for the QLW model $(b\neq 1)$ for $r=\frac{1}{2},1,$ two values of the
Weierstrass' rate $b/\mathcal{A}$ and frequency rate $r_{c}=\omega _{c}/D$.
\ For the QLW there is a non-trivial effective Hamiltonian (\ref{Heff}),
then we have to specify $r_{c}$ which is the dimensionless number comparing
the Caldeira-Legget upper frequency \cite{caldeira} against the diffusion
coefficient. In (a), (d) and (c), we chose $b/\mathcal{A}=1$ then the
trimodality of the profile is easily seen. In (a) for $r_{c}=\frac{1}{2}$
the non-diffusive behavior is more visible than in (c)\ for $r_{c}=1$, this
trimodality behavior is enhanced by increasing the Weierstrass' rate $b/%
\mathcal{A}>1$ and the frequency rate $r$. In (b) we chose $b/\mathcal{A}<1$
so the second moment of the $\mathit{P}(\Delta l,t)$ is finite, see (\ref%
{q2det}), here the profile spreads but its non-diffusion characteristics is
clearly visible.}
\label{fig3}
\end{figure}

From (\ref{Pl}) we see that in quantum mechanic the relaxation of the
initial position is controlled by a double integral: 
\begin{equation*}
\left\langle l_{0}\right\vert \rho (t)\left\vert l_{0}\right\rangle =\left( 
\frac{1}{2\pi }\right) ^{2}\int_{-\pi }^{\pi }dk_{1}\int_{-\pi }^{\pi
}dk_{2}\exp \left( \mathcal{F}\left( k_{1},k_{2},b,\mathcal{A}\right)
t\right) .
\end{equation*}

For the case $b=1$, i.e., the QRW, we can calculate analytically the
long-time asymptotic behavior in the following way%
\begin{eqnarray}
\left\langle l_{0}\right\vert \rho (t)\left\vert l_{0}\right\rangle
&=&\left( \frac{1}{2\pi }\right) ^{2}\int_{-\pi }^{\pi }dk_{1}\int_{-\pi
}^{\pi }dk_{2}\exp \left[ \frac{-i}{\hbar }\left( \mathrm{E}_{k_{1}}-\mathrm{%
E}_{k_{2}}\right) +\frac{\pi \alpha }{\beta \hbar }\left( \cos \left(
k_{1}-k_{2}\right) -1\right) \right] t  \label{t-INF} \\
&=&\left( \frac{1}{2\pi }\right) ^{2}\int_{-\pi }^{\pi }dk_{1}\int_{-\pi
}^{\pi }dk_{2}\exp \left[ \frac{i\Omega }{\hbar }\left( \cos k_{1}-\cos
k_{2}\right) +\frac{\pi \alpha }{\beta \hbar }\left( \cos \left(
k_{1}-k_{2}\right) -1\right) \right] t  \notag \\
&=&\left( \frac{1}{2\pi }\right) ^{2}\int_{-\pi }^{\pi }e^{\frac{i\Omega }{%
\hbar }t\cos k_{1}}\ dk_{1}\int_{-\pi }^{\pi }e^{\frac{-i\Omega }{\hbar }%
t\cos k_{2}}\exp \left[ \frac{\pi \alpha }{\beta \hbar }\left( \cos \left(
k_{1}-k_{2}\right) -1\right) \right] t\ dk_{2}.  \notag
\end{eqnarray}%
In the limit of $t\rightarrow \infty $ we can use the method of steepest
descent \cite{MSP}, so defining the parameter $r=\left. \frac{\Omega }{\hbar 
}\right/ \frac{\pi \alpha }{2\beta \hbar }$ it is possible to see that for $%
r\neq 0$ each Fourier integral introduces a factor $1/\sqrt{rt}$, thus the
relaxation (in one dimension) goes like $\propto 1/rt$. On the other hand,
if $r=0$ (infinite temperature limit) it is also possible to see
analytically that the dominant contribution in the double integral comes
from a small area near $k_{1}=k_{2}$, then we get for the quantum asymptotic
behavior: $\propto 1/\sqrt{t}$, like in a classic random walk (see appendix
E).

In general for $b\neq 1$ another alternative to study the quantum
decoherence, is to evaluate some correlation function associated to the
interference measurement phenomena. This can be done by analyzing the
localization probability $\left\langle l_{0}\right\vert \rho (t)\left\vert
l_{0}\right\rangle $. From the general expression for the density matrix, $%
\rho (t)$, in an arbitrary basis $\left\vert \phi _{j}\right\rangle $, we
have 
\begin{equation*}
\rho (t)=\sum_{i\neq j}C_{i}(t)C_{j}^{\ast }(t)\ \left\vert \phi
_{i}\right\rangle \left\langle \phi _{j}\right\vert +\sum_{j}\left\vert
C_{j}(t)\right\vert ^{2}\ \left\vert \phi _{j}\right\rangle \left\langle
\phi _{j}\right\vert ,
\end{equation*}%
then we may conclude that $\sum_{j}\left\vert C_{j}(t)\right\vert ^{2}\
\left\vert \left\langle \phi _{j}\right\vert \left. l_{0}\right\rangle
\right\vert ^{2}$ represents the classical probability mixture to measure
the localization at time $t$. We can take into account the relaxation of the
quantum interference, from the pure state $\rho (0)=\left\vert
l_{0}\right\rangle $ $\left\langle l_{0}\right\vert $, as follows: the
probability to measure at time $t$ the particle at the initial position is
from (\ref{Pl}):%
\begin{eqnarray*}
\left\langle l_{0}\right\vert \rho (t)\left\vert l_{0}\right\rangle
&=&\left( \frac{1}{2\pi }\right) ^{2}\int_{-\pi }^{\pi }dk_{1}\int_{-\pi
}^{\pi }dk_{2}\left[ 1-\delta \left( k_{1}-k_{2}\right) +\delta \left(
k_{1}-k_{2}\right) \right] \exp \left( \mathcal{F}\left( k_{1},k_{2},b,%
\mathcal{A}\right) t\right) \\
&=&\left( \frac{1}{2\pi }\right) ^{2}\int_{-\pi }^{\pi }dk_{1}\int_{-\pi
}^{\pi }dk_{2}\left[ 1-\delta \left( k_{1}-k_{2}\right) \right] \exp \left( 
\mathcal{F}\left( k_{1},k_{2},b,\mathcal{A}\right) t\right) \\
&&+\left( \frac{1}{2\pi }\right) ^{2}\int_{-\pi }^{\pi }dk_{1}\exp \left( 
\mathcal{F}\left( k_{1},k_{1},b,\mathcal{A}\right) t\right) .
\end{eqnarray*}%
Thus using that $\exp \left( \mathcal{F}\left( k_{1},k_{1},b,\mathcal{A}%
\right) t\right) =1$, we see that the quantum interference is characterized
by the\ localized correlation function $\chi (t)=\left\langle
l_{0}\right\vert \rho (t)\left\vert l_{0}\right\rangle -\frac{1}{2\pi },$
i.e., 
\begin{equation}
\chi (t)=\left( \frac{1}{2\pi }\right) ^{2}\int_{-\pi }^{\pi
}dk_{1}\int_{-\pi }^{\pi }dk_{2}\exp \left( \mathcal{F}\left( k_{1},k_{2},b,%
\mathcal{A}\right) t\right) -\frac{1}{2\pi }.  \label{X(t)}
\end{equation}%
In figure 4 we have plotted the localized correlation $\chi (t)$ as a
function of time for different values of Weierstrass' parameters $b,\mathcal{%
A}$, and for several frequency rates $r,r_{c}$ characterizing different
energy regimes in the QME. In this figure the long-time coherence of $\chi
(t)$ for the QLW can be compared against the diffusive decoherence
corresponding to the QRW case, i.e., $\propto 1/t$. On the other hand, only
in the infinite temperature limit $r\rightarrow 0$ the classical behavior $%
\propto 1/\sqrt{t}$ is reobtained (see Appendix E). There is some numerical
evidence that $\chi (t)$ for the QLW is in fact bounded from above $\propto
1/\sqrt{t}$, and its long-time wavy behavior is the signature of the slow
decoherence due to the scale-free hopping structure.

\begin{figure}[t!]
\includegraphics[width=0.95 \columnwidth,clip]{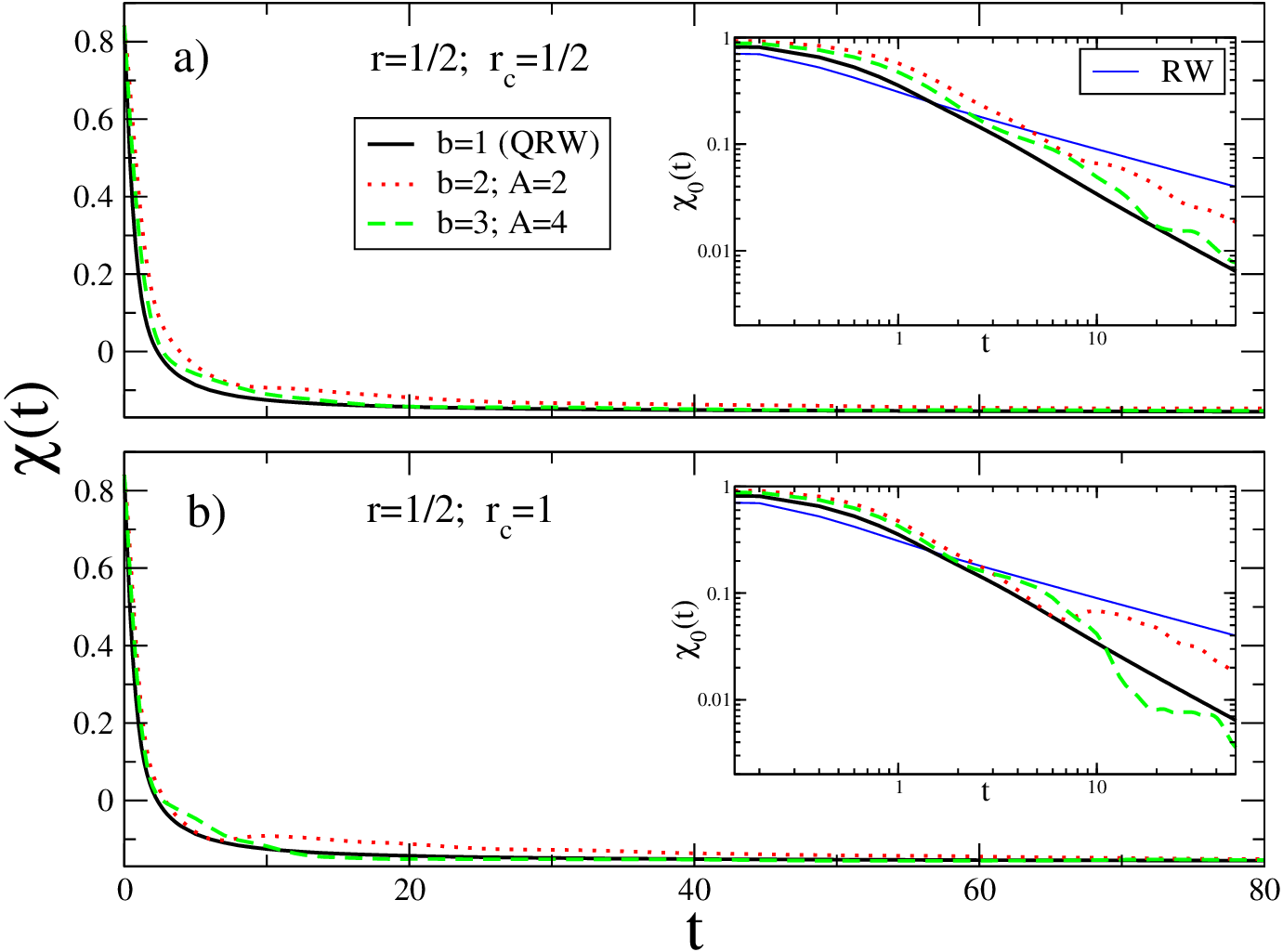}
\caption{Plot of the localized correlation $\chi (t)$ for the QLW as a
function of dimensionless time $t$ when the system is prepared in the pure
state $\rho (0)=\left\vert l_{0}\right\rangle \left\langle l_{0}\right\vert $%
. The function $\chi (t)$ is plotted with $r=\frac{1}{2}$ for two values of
Weierstrass' rate $b/\mathcal{A}$ and frequency rate $r_{c}$. The insets
show the log-log asymptotic behavior of $\chi _{0}(t)\equiv \chi (t)+\frac{1%
}{2\pi }$ against the QRW decay $\propto 1/t$ (wide straight line), also in
the inset we have plotted (thin straight line) the classical RW localized
probability $P_{0}(t)\propto 1/\sqrt{t}$, see (\ref{Pt}). The long-time
coherent behavior of $\chi (t)$ for the QLW is much more visible for large
values of $r_{c}$. Note that if $r\neq 0$ none of the cases: QLW or the QRW
behave asymptotically like a \textit{classical} 1D walk (see appendix E for
details). The localized quantum correlation $\chi (t)$ is bounded from above 
$\propto 1/\sqrt{t}$ and its wavy behavior for long-time (shown in the
inset) is the signature of the slow decoherence due to Levy's hopping
structure.}
\label{fig4}
\end{figure}

In order to study the decoherence of a pure state due to the interaction
with the thermal bath $B$, we propose here to analyze the time evolution of $%
Tr\left[ \rho ^{2}\right] $, this function is some time called the quantum
purity. Assuming that at time $t=0$ the system was prepared in the Wannier
pure state $\rho (0)=\left\vert l_{0}\right\rangle $ $\left\langle
l_{0}\right\vert $, the quantum purity is given by%
\begin{eqnarray*}
Tr\left[ \rho (t)^{2}\right] &=&\int_{-\pi }^{\pi }dk_{1}\int_{-\pi }^{\pi
}dk_{2}\left\langle k_{1}\right\vert \rho (t)\left\vert k_{2}\right\rangle
\left\langle k_{2}\right\vert \rho (t)\left\vert k_{1}\right\rangle \\
&=&\left( \frac{1}{2\pi }\right) ^{2}\int_{-\pi }^{\pi }dk_{1}\int_{-\pi
}^{\pi }dk_{2}\exp \left[ \mathcal{F}\left( k_{1},k_{2},b,\mathcal{A}\right)
+\ \exp \mathcal{F}\left( k_{2},k_{1},b,\mathcal{A}\right) \right] t.
\end{eqnarray*}%
Introducing the definition of the rates $r,r_{c}$ in the explicit expression
of $F\left( k_{1},k_{2},b,\mathcal{A}\right) $, from (\ref{evo3}) we can see
that 
\begin{equation}
Tr\left[ \rho (t)^{2}\right] =\left( \frac{1}{2\pi }\right) ^{2}\int_{-\pi
}^{\pi }dk_{1}\int_{-\pi }^{\pi }dk_{2}\ \exp \left\{ \left. \mathcal{F}%
\left( k_{1},k_{2},b,\mathcal{A}\right) \right\vert _{r=r_{c}=0}\
2Dt\right\} .  \label{purity}
\end{equation}%
Thus the time can be rescaled in unit of the diffusion constant $D$, we also
see that the behavior of the purity will be similar to the\ localized
correlation function $\chi (t)$ but subtracting the quantum time-dependent
oscillations, this is so because $F\left( k_{1},k_{2},b,\mathcal{A}\right) $
is evaluated at $r=r_{c}=0$, compare with (\ref{X(t)}). In figure 5 we show $%
Tr\left[ \rho (t)^{2}\right] $ in a log-log plot for different values of
Weierstrass' parameters $b,A$. From this plot the non-trivial long tail
behavior of the QLW can clearly be compared against the QRW case. For the
case $b=1$ (the NN walks) and using the method of the steepest descent \cite%
{MSP} from (\ref{purity}) is it possible to prove that the asymptotic
behavior of the purity for the QRW is $\propto 1/\sqrt{Dt}$, this asymptotic
regime is also shown (straight line) in figure 5. For the QLW $(b>1)$ we can
see that the asymptotic behavior looks like $\propto 1/t^{\xi }$, where the
exponent $\xi $ is a non-trivial function of Weierstrass' parameters $b,A$.
In the inset of this figure we also show the long-tail exponent $\xi $
versus the Weierstrass parameter $A$, here a transition in the behavior of $%
\xi $ for $b/A\lessgtr 1$ can be seen. From this inset it is also possible
to see that if $A\gg 1$ the exponent $\xi $ goes to $0.5$ corresponding to
the QRW case.
\begin{figure}[t!]
\includegraphics[width=0.95 \columnwidth,clip]{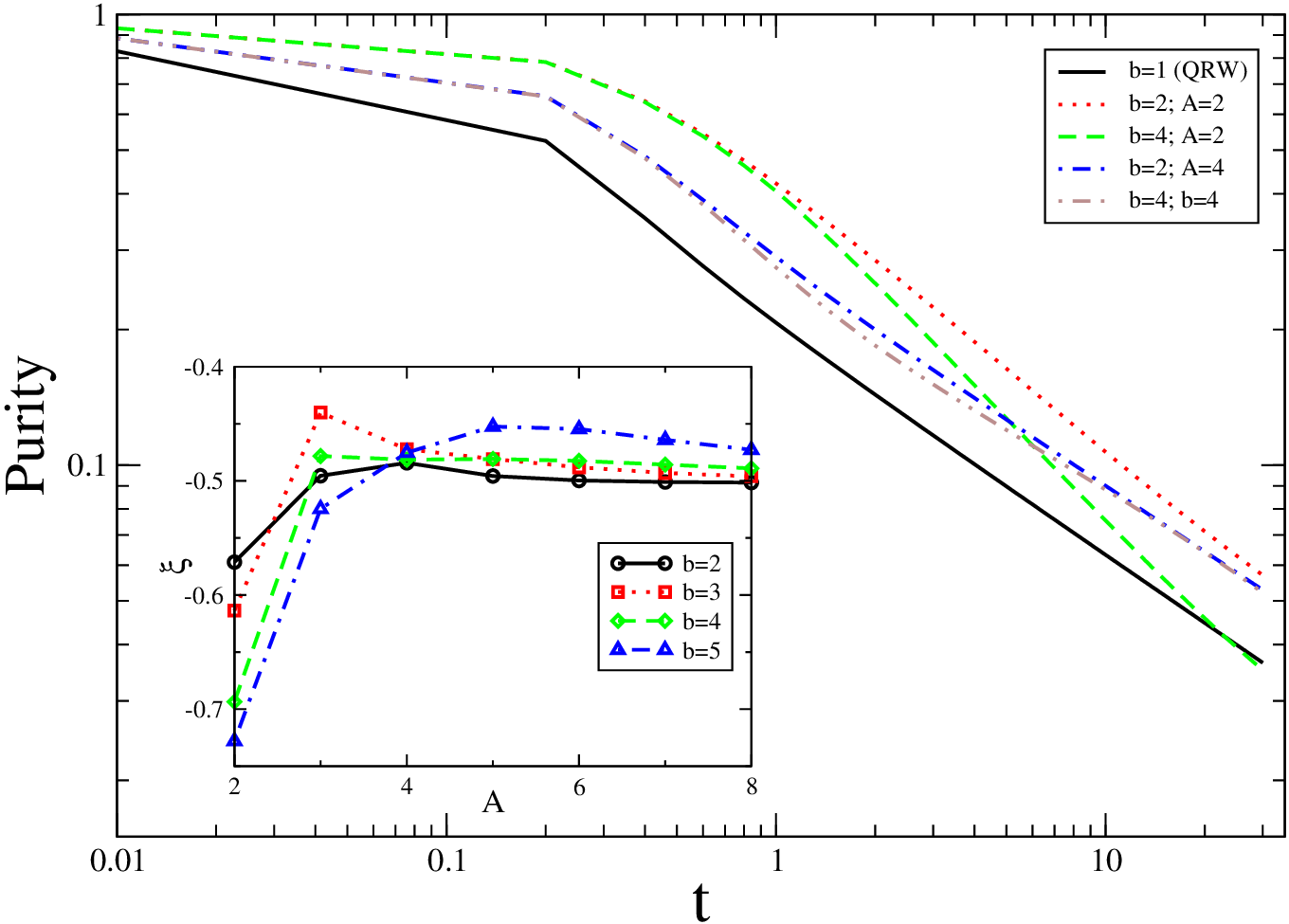}
\caption{Log-log plot of the quantum purity $Tr\left[ \rho ^{2}\right] $ as a
function of dimensionless time $t$ when the system is prepared in the pure
state $\rho (0)=\left\vert l_{0}\right\rangle \left\langle l_{0}\right\vert $%
. The purity function is plotted for several values of Weierstrass' rate $b/%
\mathcal{A}$. The purity for the QRW (straight line) shows the predicted
long-time decoherent decay $\propto 1/\sqrt{t}$, see (\ref{purity}). The
long-time asymptotic behavior for the QLW shows an asymptotic behavior like $%
\propto 1/t^{\xi }$ where the exponent $\xi $ depends on the Weierstrass'
parameters $b,\mathcal{A}$ in a non-trivial way, see inset.}
\label{fig5}
\end{figure}

\subsubsection{Quantum decoherence from a coherent preparation in $%
\left\vert k\right\rangle $}

Now we will study the relaxation from a coherente preparation in the Fourier
bases. In order to simplify this analysis we adopt here the following
initial preparation for the system%
\begin{equation}
\rho (0)=\frac{1}{k_{c}}\int_{0}^{k_{c}}dk\int_{0}^{k_{c}}dk^{\prime
}\left\vert k\right\rangle \left\langle k^{\prime }\right\vert ,  \label{M1}
\end{equation}%
where $k_{c}\leq \pi $ characterizes the \textit{initial} uniform
probability distribution in Fourier space. Therefore the mean-value of the
Fourier modes appearing in the initial preparation is $\int_{-\pi }^{\pi
}k\rho _{kk}(0)dk=k_{c}/2$. On the other hand, from (\ref{M1}) we get%
\begin{eqnarray}
\rho _{k_{1}k_{2}}(0) &=&\left\langle k_{1}\right\vert \rho (0)\left\vert
k_{2}\right\rangle =\frac{1}{k_{c}}\int_{0}^{k_{c}}dk\int_{0}^{k_{c}}dk^{%
\prime }\delta \left( k_{1}-k\right) \delta \left( k^{\prime }-k_{2}\right)
\label{M2} \\
&=&\frac{1}{k_{c}}\Theta \left( k_{c}-k_{1}\right) \Theta \left(
k_{1}\right) \Theta \left( k_{c}-k_{2}\right) \Theta \left( k_{2}\right) , 
\notag
\end{eqnarray}%
where $\Theta \left( z\right) $ is the step function. As indicated before
the uniform initial probability distribution will be invariant in time $\rho
_{kk}(0)=\frac{1}{k_{c}}\Theta \left( k_{c}-k\right) \Theta \left( k\right) $%
, and fulfills normalization:%
\begin{equation*}
\int_{-\pi }^{\pi }dk\rho _{kk}(0)=\frac{1}{k_{c}}\int_{-\pi }^{\pi }\Theta
\left( k_{c}-k\right) \Theta \left( k\right) dk=1.
\end{equation*}%
Therefore the probability $\left\langle l\right\vert \rho (t)\left\vert
l\right\rangle $ will have a time dependent spreading, but with a well
defined mean-velocity characterized by the Fourier value $k_{c}$. Using (\ref%
{M2}) as the initial preparation of the system, we get for the
time-dependent probability to be at lattice site $l$ 
\begin{eqnarray}
\left\langle l\right\vert \rho (t)\left\vert l\right\rangle &=&\int_{-\pi
}^{\pi }dk_{1}\int_{-\pi }^{\pi }dk_{2}\left\langle l\right. \left\vert
k_{1}\right\rangle \left\langle k_{1}\right\vert \rho (t)\left\vert
k_{2}\right\rangle \left\langle k_{2}\right\vert \left. l\right\rangle 
\notag \\
&=&\frac{1}{2\pi }\int_{-\pi }^{\pi }dk_{1}\int_{-\pi }^{\pi }dk_{2}\exp %
\left[ i\left( k_{1}-k_{2}\right) l\right] \left\langle k_{1}\right\vert
\rho (t)\left\vert k_{2}\right\rangle  \notag \\
&=&\frac{1}{2\pi k_{c}}\int_{0}^{k_{c}}dk_{1}\int_{0}^{k_{c}}dk_{2}\exp %
\left[ i\left( k_{1}-k_{2}\right) l\right] \exp \left( \mathcal{F}\left(
k_{1},k_{2},b,\mathcal{A}\right) t\right) .  \label{mixture}
\end{eqnarray}%
Note that at $t=0$ the initial distribution was delocalized in all the
lattice according to: 
\begin{eqnarray}
\left\langle l\right\vert \rho (0)\left\vert l\right\rangle &=&\frac{1}{2\pi
k_{c}}\int_{0}^{k_{c}}dk_{1}\int_{0}^{k_{c}}dk_{2}\exp \left[ i\left(
k_{1}-k_{2}\right) l\right]  \label{mixt0} \\
&=&\frac{1}{\pi k_{c}l^{2}}\left( 1-\cos k_{c}l\right) ,\qquad l\in \left(
-\infty ,+\infty \right) .  \notag
\end{eqnarray}%
In figure 6 we have plotted the probability $\left\langle l\right\vert \rho
(t)\left\vert l\right\rangle $ from (\ref{mixture}) as a function of the
position $l$ for different values of time. The non-diffusive behavior as
well as the reentrance phenomenon of the (driven) profile probability $%
\left\langle l\right\vert \rho (t)\left\vert l\right\rangle $ for the QLW
can clearly be seen. What we call\textit{\ reentrance phenomenon} is just
the cooperative result of the clustering that occurs for $b>A$ \ and the
quantum coherence. Classically the clustering is certain when $0<\mu <1$;
only in the case $\mu \geq 1$ the walker ultimately returns to fill in any
gaps in the set of sites visited and the clustering eventually disappears 
\cite{HMSch}.
\begin{figure}[t!]
\includegraphics[width=0.95 \columnwidth,clip]{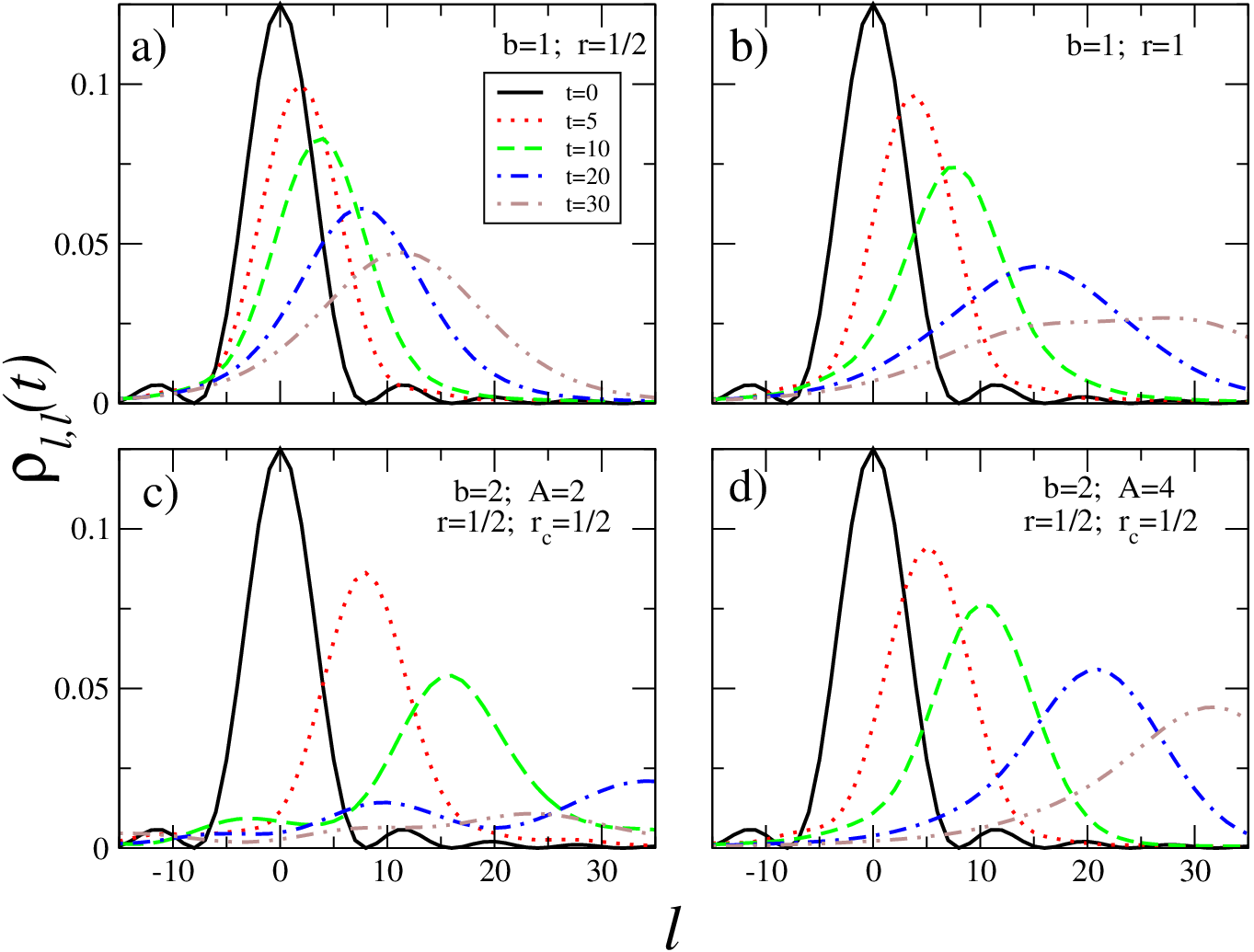}
\caption{Plot of the probability $\rho _{l,l}(t)\equiv \left\langle
l\right\vert \rho (t)\left\vert l\right\rangle $ as a function of lattice
site $l$ for several values of dimensionless time $t$, Weierstrass' rate $b/%
\mathcal{A}$, frequency rate $r$ and for $r_{c}=\frac{1}{2}$, when the
system is prepared in a Fourier coherent state like in (\ref{M1}). The
straight line corresponds to the profile at $t=0$ as is given in (\ref{mixt0}%
) for $k_{c}=\pi /4$. figures (a) and (b) correspond to the QRW $(b=1)$ for
two values of rate $r$, the profile moves to the right with a thermal mean
velocity as predicted in (\ref{MVoo}). In figure (c) it is possible to see
the "reentrance" of the probability to small values of $l$ for increasing
time $(t=20)$, this phenomenon is due to the \textit{clustering} of the
visited sites and occurs above the critical value $b/\mathcal{A}>1.$}
\label{fig6}
\end{figure}

The agreement between the driven profile and the prediction associated to
the pseudo-momentum is clearly visible from figure 6. Note that for the
coherent preparation (\ref{M1}), the thermal mean value of the
pseudo-momentum will be different from zero 
\begin{eqnarray}
\left\langle \mathbf{p}(t)\right\rangle &=&Tr\left[ \mathbf{p}\rho (t)\right]
=\int_{-\pi }^{\pi }dk\int_{-\pi }^{\pi }dk_{1}\left\langle k\right\vert 
\mathbf{p}\left\vert k_{1}\right\rangle \left\langle k_{1}\right\vert \rho
(t)\left\vert k\right\rangle  \notag \\
&=&\int_{-\pi }^{\pi }dk\int_{-\pi }^{\pi }dk_{1}\ \mathbf{p}_{k}\delta
\left( k_{1}-k\right) \left\langle k_{1}\right\vert \rho (t)\left\vert
k\right\rangle  \notag \\
&=&\frac{1}{k_{c}}\int_{0}^{k_{c}}dk\ \mathbf{p}_{k}  \notag \\
&=&\frac{m\Omega }{\hbar k_{c}}\left( \frac{\mathcal{A-}1}{\mathcal{A}}%
\right) \sum_{n}\frac{1}{\mathcal{A}^{n}}\left[ 1-\cos \left(
k_{c}b^{n}\right) \right] ,  \label{MVoo}
\end{eqnarray}%
where we have used (\ref{momento}). Interestingly taking $b=1$ in (\ref{MVoo}%
) we get that for the coherent initial preparation (\ref{M1}) the QRW has a
thermal mean velocity characterized by $\left\langle \mathbf{p}%
(t)\right\rangle _{\text{QRW}}/m=\frac{\Omega }{\hbar k_{c}}\left[ 1-\cos
\left( k_{c}\right) \right] .$Therefore we may conclude that for the initial
preparation (\ref{M1}), the driven profile of the QLW moves faster than the
corresponding profile for the QRW, as can be seen from the following
inequality:

\begin{equation*}
\frac{\hbar k_{c}}{m\Omega }\left\langle \mathbf{p}(t)\right\rangle _{\text{%
QLW}}=\left( \frac{\mathcal{A-}1}{\mathcal{A}}\right) \sum_{n=0}\frac{1}{%
\mathcal{A}^{n}}\left[ 1-\cos \left( k_{c}b^{n}\right) \right] >\frac{\hbar
k_{c}}{m\Omega }\left\langle \mathbf{p}(t)\right\rangle _{\text{QRW}}
\end{equation*}

Note from (\ref{MVoo}) that even when the eigenvalue $\mathbf{p}_{k}$, see (%
\ref{momento}), is not defined for $b>\mathcal{A}$, the double Fourier
integral restores the regularity, i.e., the thermal mean-value of the
pseudo-momentum $\left\langle \mathbf{p}(t)\right\rangle $ is well defined
for any value of the Weierstrass parameters $b,\mathcal{A}$. Of course this
is nothing more than the fact that by using the initial preparation (\ref{M1}%
) the probabilistic profile move to the right with a finite velocity. In the
case $\mathcal{A}\gg 1$ the value of the thermal\ mean velocity goes to the
one corresponding to the NN model.

\section{Discussions}

In this work we have introduced the Weierstrass shift operators to build up
the quantum master equation (also called Born-Markov equation) governing the
quantum Levy walk process. We have started from the microscopic dynamics of
a free particle, in a lattice, in interaction with a thermal phonon bath.
The coupling with the heat bath was chosen proportional to shift operators
(see (\ref{Vi}) and (\ref{Bi})), but any type other coupling can be worked
out using the recipes that we have write down in appendices A, B, C. Then
the Kossakowski-Lindblad generator of the quantum semigroup, which is the
starting point to analyze the positivity condition of the structure matrix,
was built up. For the present model it was not necessary to apply Davies's
formalism (random phase approximation) to calculate the generator because
Weierstrass' shift operators, in the Heisenberg representation, are constant
in time. This results in that the\ complete positivity condition of the
generator was assured. The quantum Levy walk Hamiltonian shows for $b>%
\mathcal{A}$ that the DOS has a complex null-set inside the Brillouin zone,
because the continuous eigenenergy $\mathcal{E}_{k}(b,\mathcal{A})$ is
non-differentiable when $b/\mathcal{A}>1$. We proved that several physical
objects show the signature of this singular characteristic. A rigorous
calculus based on the scaling properties of the eigenenergy (\ref{scaling})
leads to the conclusion that the record of $\mathcal{E}_{k}(b,\mathcal{A})$
as a function of $k$ shows a critical behavior for values $b>A$. In fact,
the eigenenergy $\mathcal{E}_{k}(b,\mathcal{A})$ is self-affine with a
(Box-counting) fractal dimension $D=2-\mu ,$ for $0<\mu \equiv \ln \mathcal{A%
}/\ln b<1$.

The analytical solution of the quantum master equation has been found in the
Fourier basis, see (\ref{evo2}). In particular the evolution equation for
the second moment (\ref{q2}), and its solution (\ref{q2det}), has been
calculated. Thus we prove that the thermal second moment is not divergent
under the restriction $b<\mathcal{A}$, which is different when compared with
the classical Weierstrass universality $b^{2}<\mathcal{A}$. The present
quantum master equation can also be used to get information about higher
moments and correlations of the quantum Levy walk process. The quantum
decoherence has been characterized analyzing the probability to be in a
lattice site considering different initial preparations for the system,
i.e.,\ we have worked out two different preparations for the reduced density
matrix: a pure state in the Wannier basis, and a coherent preparation in the
reciprocal Fourier basis. We have also examined the interference correlation
function associated to the localization measurement process. In the present
paper we work in a discrete infinite dimensional Hilbert space $\mathbf{H}$
with lattice parameter $\epsilon =1$. The important issue about discrete
against continuum dissipative model can also be tackled, for the present
scale free Hamiltonian, by introducing the small lattice parameter limit $%
\epsilon \rightarrow 0$ and using the Wigner transformation in the quantum
master equation (\ref{QMEa}) (see end of appendix B for details).

The continuous eigenvalues of the scale free Hamiltonian (\ref{H}), and the
effective Hamiltonian (\ref{Heff}) have been studied as a function of the
Fourier number $k$ (the \textit{tight-binding} case corresponds to take $b=1$
in (\ref{Ek})). The quantum Levy walk Hamiltonian ($H_{S}$)\ for different
values of Weierstrass' parameters $b$ and $\mathcal{A}$ has been plotted
showing the non-differentiable structure of its spectrum $\mathcal{E}_{k}(b,%
\mathcal{A)}$ when the rate $b/\mathcal{A}$ is larger than $1$, this was
also quoted in connection to the critical behavior of the DOS (\ref{DOS}).
We show that the continuous eigenvalue of the effective Hamiltonian ($%
H_{S}-\hbar \omega _{c}\ a^{\dagger }a$) has fluctuations of larger
amplitude for increasing energy rate $r_{e}=\hbar \omega _{c}/\Omega $, see
figure 2.

For the transport of a quantum Levy particle the non-trivial effective
Hamiltonian can be characterized in terms of the frequency rate $%
r_{c}=\omega _{c}/D$, this dimensionless number compares the Caldeira-Legget
upper frequency \cite{caldeira} against the diffusion coefficient $D=\pi
\alpha /2\beta \hbar $, being $\alpha $ the dissipative parameter and $\beta 
$ the inverse of the temperature of the bath (see \ref{F(a)}). When the
system is prepared in a pure Wannier state, the probability to be at
distance $\Delta l$ from the initial position $l_{0}$, i.e., $\mathit{P}%
(\Delta l,t)$, was studied as a function of $\Delta l$ for several values of
time $t$ and for several values of the frequency rates $r=\frac{\Omega }{%
\hbar }/D,\ r_{c}.$ The larger the frequency rate $r$ is the longer the
quantum coherence persists before the profile starts to spread. The opposite
case $r\rightarrow 0$ corresponds to the high temperature limit. We have
shown that the profile $\mathit{P}(\Delta l,t)$ spreads in a non-diffusive
way when $b\neq 1$, i.e., characterizing a Levy-like behavior. The larger
the rate $r_{c}$ is the more the time oscillations of $\mathit{P}(\Delta
l=0,t)$ there are. We have also shown that the profile shows an important 
\textit{trimodality}, this behavior is intensified by increasing
Weierstrass' rate $b/\mathcal{A}$ above the critical value $b/\mathcal{A}=1.$
The localized correlation $\chi (t)$ has been studied as a function of time $%
t$ when the Levy particle is prepared in the pure state $\rho (0)=\left\vert
l_{0}\right\rangle \left\langle l_{0}\right\vert $. The function $\chi (t)$
has been plotted for different values of Weierstrass' parameters $b,\mathcal{%
A}$ and frequency rate $r_{c}$, we have shown that the asymptotic decay of $%
\chi (t)$ does not have the 1D classical $\propto 1/\sqrt{t}$ diffusive
behavior, on the contrary the localized correlation function $\chi (t)$
shows a long-time coherent persistence, see figure 4. The quantum purity $Tr%
\left[ \rho ^{2}\right] $\ has been studied as a function of time $t$ for
several values of Weierstrass' parameters $b,\mathcal{A}$. The purity for
the quantum random walk $(b=1)$ shows the predicted long-time decay $\propto
1/\sqrt{t}$ for the NN case (\ref{purity}). Nevertheless, for the quantum
Levy particle the purity has a long-time asymptotic behavior that looks like 
$\propto 1/t^{\xi }$, where the exponent $\xi $ depends on the Weierstrass
parameters $b/\mathcal{A}$ in a non-trivial way, see inset of figure 5.

When the system is prepared in a coherent state as in (\ref{M1}), the
probability $\left\langle l\right\vert \rho (t)\left\vert l\right\rangle $
for the quantum Levy particle has been studied as a function of $l$ for
several values of time $t$, Weierstrass' parameters $b,\mathcal{A}$ and
frequency rate $r$. We have checked numerically that for the coherent
preparation (\ref{M1}), with $k_{c}=\pi /4$, the profile moves to the right
with a thermal mean-velocity as predicted in (\ref{MVoo}). We have also
shown that the profile has an interesting \textit{reentrance} behavior. In
figure 6c the reentrance of the probability to small values of $l$ for
increasing time is clearly visible, this phenomenon is the result of the
cooperative phenomena of\textit{\ clustering} (for the critical value $b>A$)
and \textit{quantum coherence}. The trimodality that occurs from a pure
initial preparation like $\rho (0)=\left\vert l_{0}\right\rangle
\left\langle l_{0}\right\vert $, also is shown to occur when there is
clustering in the walks.

In the present framework it is also possible to analyze the quantum jump
picture (\ref{irre2}) which is of great value to measure the dissipative
decoherence \cite{kossa,dalibard}. This last analysis can be done by
studying the fluctuation superoperator $F\left[ \bullet \right] $ in the
Fourier basis, i.e., (\ref{F(a)}). We have prove in the appendix C, that the
superoperator $F\left[ \bullet \right] $ can be handled in terms of the
elements of the operators $a,a^{\dag }$, which for $b\neq 1$ have a
non-trivial Fourier structure resembling the characteristic function of the
classical Weierstrass random walk. Further works along these lines are in
progress.

In conclusion, the open quantum Levy lattice model introduced here for
transport in a infinite dimensional Hilbert space, is of remarkable
usefulness for calculating analytically many interesting measures of
irreversibility. The important issue of decoherence in a dissipative long
range model was tackled analytically. The analysis of other irreversible
measures like entropy in the context of the coherence-vector formulation,
etc. \cite{alicki}, can also be worked out in the present framework and is
in progress.

\section*{Acknowledgments}

M.O.C. thanks A.K. Chattah for early discussions on the QRW model, Prof. V.
Gr\"{u}nfeld for the English revision of the manuscript, and grants from
SECTyP, Universidad Nacional de Cuyo, Argentina. M.N. thanks the fellowship
from CONICET.

\appendix

\section{Quantum dynamic semigroups revisited}

Quantum dynamic semigroups are the generalization of Markov semigroups for
non-commutative algebras \cite{spohn,alicki}. In the Markovian
approximation, Kossakowski and Lindblad established the form of the Quantum
Master Equation (QME) in order that the evolution of the system of interest
shall correspond to a quantum dynamic semigroup (these semigroups are also
called \textquotedblleft completely positive semigroups\textquotedblright\
CPS). In the \textit{structural theorem}, Lindblad \cite{lindblad}
established that the \textit{generator} of a CPS, acting on the reduced
density matrix of the system, $\rho $, has the form: $L\left[ \bullet \right]
\equiv -\frac{i}{\hbar }[H_{eff},\bullet ]+\frac{1}{2}\sum_{\alpha }\left[
V_{\alpha }\bullet ,V_{\alpha }^{\dagger }\right] +\left[ V_{\alpha
},\bullet V_{\alpha }^{\dagger }\right] $, where $V_{\alpha }$ are bounded
operators. If the system has a finite number of degrees of freedom, the
expression for the generator $L\left[ \bullet \right] $ can alternatively be
written in a different way. Consider for example the algebra of the $N\times
N$ complex matrices, then assume that the set of operators $\{V_{\alpha
}\}_{\alpha =1}^{N^{2}-1}$ fulfilling $Tr[V_{\alpha }^{\dagger }V_{\alpha
^{\prime }}]=\delta _{\alpha ,\alpha ^{\prime }},$ is a basis in that space
(i.e.: the basis is orthonormal with respect to the \textit{scalar product}: 
$<A|B>=Tr[A^{\dagger }B]$). In term of this basis, the generator of the CPS
is \cite{kossa}

\begin{equation}
L\left[ \bullet \right] =-\frac{i}{\hbar }[H_{eff},\bullet ]+\frac{1}{2}%
\sum_{\alpha ,\gamma =1}^{N^{2}-1}a_{\alpha \gamma }(\left[ V_{\alpha
}\bullet ,V_{\gamma }^{\dagger }\right] +\left[ V_{\alpha },\bullet
V_{\gamma }^{\dagger }\right] ),  \label{eq2}
\end{equation}%
where the matrix of elements $a_{\alpha \gamma }$ is positive-definite. This
generator is written in the Schr\"{o}dinger representation and acts on the
density matrix of the system of interest. In the Heisenberg representation
the dual generator $L^{\ast }\left[ \bullet \right] ,$ defined as $%
Tr[L^{\ast }[A]\rho ]=Tr[AL[\rho ]]$ acts on any physical observable $A$
(Hermitian operator). Now we define the superoperator 
\begin{equation}
F[\bullet ]=\sum_{\alpha ,\gamma =1}^{N^{2}-1}a_{\alpha \gamma }V_{\alpha
}\bullet V_{\gamma }^{\dagger },  \label{F-fin}
\end{equation}%
and we consider its dual $F^{\ast }[\mathbf{\bullet }]$ evaluated in the
identity operator $\mathbf{1}$, i.e., $F^{\ast }[\mathbf{1}]=\sum_{\alpha
,\gamma =1}^{N^{2}-1}a_{\alpha \gamma }V_{\gamma }^{\dagger }V_{\alpha }$ .
Then the generator (\ref{eq2}) can be written in the compact form 
\begin{equation}
L\left[ \bullet \right] =-\frac{i}{\hbar }[H_{eff},\bullet ]+F[\bullet ]-%
\frac{1}{2}\left\{ F^{\ast }[\mathbf{1}],\bullet \right\} _{+},  \label{eq2b}
\end{equation}%
\bigskip The operator $\frac{1}{2}F^{\ast }[\mathbf{1}]$ can be regarded as
a dissipative operator, and $F[\bullet ]$ the fluctuating superoperator.
Equations (\ref{F-fin}) and (\ref{eq2b}) allow us to define the \textit{%
structure matrix} $[a_{\alpha \gamma }]$. This matrix contain the
information concerning the relaxation times of the dynamic system in contact
with a thermal bath. We say that a generator with the structure (\ref{eq2b}%
), with an Hermitian matrix $[a_{\alpha \gamma }]$ has the \textit{form }of
a Kossakowski-Lindblad generator (KL) \cite{jpa}. We note that the
corresponding semigroup is completely positive \textit{if and only if} 
\textit{\ }$[a_{\alpha \gamma }]$ is a positive-definite matrix, and it is
equivalent to say that the generator $L\left[ \bullet \right] $ fulfills the
structural theorem. Alternatively, we will say that when \textit{\ }$%
[a_{\alpha \gamma }]\geq 0$ the generator is a \textit{well defined} KL
generator \cite{sandu,dekker}.

The formal solution of the QME, $\dot{\rho}=L[\rho ]$, is 
\begin{equation}
\rho (t)=\exp \left\{ (-\frac{i}{\hbar }[H_{eff},\bullet ]+F[\bullet ]-\frac{%
1}{2}\left\{ F^{\ast }[\mathbf{1}],\bullet \right\} _{+})\,t\right\} \rho
(0).  \label{irre}
\end{equation}%
This expression can be put in the form \cite{plenio} 
\begin{eqnarray}
\rho (t) &=&\sum_{m=0}^{\infty
}\int_{0}^{t}dt_{m}\int_{0}^{t_{m}}dt_{m-1}\cdots \int_{0}^{t_{2}}dt_{1}
\label{irre2} \\
&&\times \{S(t-t_{m})F[\bullet ]S(t_{m}-t_{m-1})\cdots F[\bullet
]S(t_{1})\}\rho (0).  \notag
\end{eqnarray}%
In this way the dynamics of the system can be interpreted as if it were
composed by quantum jumps (associated to the superoperator $F[\bullet ]$)
and in between them there is a smooth non-unitary evolution determined by 
\begin{equation*}
S(t)\rho =\exp \left\{ (-\frac{i}{\hbar }[H_{eff},\bullet ]-\frac{1}{2}%
\left\{ F^{\ast }[\mathbf{1}],\bullet \right\} _{+})\,t\right\} \rho \equiv 
\mathbf{N}(t)\rho \mathbf{N}^{\dagger }(t),
\end{equation*}%
where $\mathbf{N}(t)=\exp (-\left[ \frac{i}{\hbar }H_{eff}+\frac{1}{2}%
F^{\ast }[\mathbf{1}]\right] t)$ characterizes decay. This representation is
very suitable for describing the decoherence of the off-diagonal elements of
the density matrix \cite{kossa,dalibard}.

\subsection{The quantum master equation and the second order approximation}

It is known that the QME arising from second order perturbation theory has,
in general, the KL \textit{form} (\ref{eq2b}). To see this assume that the
total Hamiltonian is of the form: $H_{T}=H_{S}+H_{B}+\theta H_{I},$ and that
the system $\mathcal{S}$ interacts with a equilibrium thermal bath $\mathcal{%
B}$ through the term $\theta H_{I}$ ($\theta $ is the coupling intensity),
also we assume that the initial condition for the total density matrix can
be written in the form $\rho _{T}(0)=\rho (0)\otimes \rho _{B}^{e}$, where $%
\rho _{B}^{e}$ is the equilibrium density matrix of the bath. Now consider
the Liouville equation for the total density matrix and trace out the bath
variables, keeping only up to the second order $\mathcal{O}(\theta ^{2})$.
This procedure gives a QME for the reduced density matrix of the system $%
\rho $ having a KL\ \textit{form} where the generator is defined through an
effective Hamiltonian $H_{eff}$ and the superoperator $F\left[ \bullet %
\right] $ \cite{jpa}, 
\begin{equation}
H_{eff}=H_{S}-i\frac{\theta ^{2}}{2\hbar }\int_{0}^{\infty }d\tau \
Tr_{B}\left( \left[ H_{I},H_{I}(-\tau )\right] \rho _{B}^{e}\right) ,
\label{eq3}
\end{equation}

\begin{equation}
F\left[ \rho (t)\right] =\left( \frac{\theta }{\hbar }\right)
^{2}\int_{0}^{\infty }d\tau \ Tr_{B}\left( H_{I}\,\,\,\rho (t)\otimes \rho
_{B}^{e}\,\,H_{I}(-\tau )+H_{I}(-\tau )\,\,\,\rho (t)\otimes \rho
_{B}^{e}\,\,H_{I}\right) .  \label{eq4}
\end{equation}%
Where $H_{I}(-\tau )=e^{-i\tau (H_{S}+H_{B})/\hbar }H_{I}\,e^{i\tau
(H_{S}+H_{B})/\hbar }$. We remark that this structure for the generator is
independent of any particular system $\mathcal{S}$ under consideration; it
is also valid for finite or infinite dimensional Hilbert spaces. Now we
consider the interaction Hamiltonian to be characterized by the direct
product of operators: 
\begin{equation}
H_{I}=\sum\limits_{\beta }V_{\beta }\otimes B_{\beta },\,  \label{hi}
\end{equation}%
Then using explicitly the Hermitian condition of $H_{I}$, $H_{eff}$ and the
form of the superoperator $F\left[ \bullet \right] $, we can write 
\begin{equation}
H_{eff}=H_{S}-i\frac{\theta ^{2}}{2\hbar }\sum\limits_{\alpha \beta
}\int_{0}^{\infty }d\tau \left( \chi _{\alpha \beta }(-\tau )V_{\alpha
}^{\dagger }V_{\beta }(-\tau )-\chi _{\alpha \beta }^{\ast }(-\tau )V_{\beta
}^{\dagger }(-\tau )V_{\alpha }\right) ,  \label{eq3c}
\end{equation}%
\begin{equation}
F\left[ \bullet \right] =\left( \frac{\theta }{\hbar }\right)
^{2}\sum\limits_{\alpha \beta }\int_{0}^{\infty }d\tau \left( \chi _{\alpha
\beta }(-\tau )V_{\beta }(-\tau )\bullet V_{\alpha }^{\dagger }+\chi
_{\alpha \beta }^{\ast }(-\tau )V_{\alpha }\bullet V_{\beta }^{\dagger
}(-\tau )\right) .  \label{eq4b}
\end{equation}%
Here we have introduced the correlation functions of the thermal bath:%
\begin{equation}
\chi _{\alpha \beta }(-\tau )\equiv Tr_{B}\left( \rho _{B}^{e}\ B_{\alpha
}^{\dagger }\,B_{\beta }(-\tau )\right) ,  \label{correla}
\end{equation}%
where%
\begin{eqnarray*}
B_{\alpha }(-\tau ) &\equiv &\exp (-i\tau H_{B}/\hbar )B_{\alpha }\exp
(i\tau H_{B}/\hbar ), \\
V_{\beta }(-\tau ) &\equiv &\exp (-i\tau H_{S}/\hbar )V_{\beta }\exp (i\tau
H_{S}/\hbar ).
\end{eqnarray*}%
Because the thermal bath is stationary the correlation function fulfills the
symmetry condition: $\chi _{\alpha \beta }(-\tau )=\chi _{\beta \alpha
}^{\ast }(\tau ).$ The KL \textit{form }(\ref{eq2b}) allow us to analyze its
possible positivity. We do this because, even when the QME up to the second
order approximation can be written in a KL \textit{form}, it is not possible
to assure that the semigroup will be completely positive.

In the case of working with a finite dimensional Hilbert space the analysis
of the structure $[a_{\alpha \gamma }]$ allows us to introduce a\textit{\
necessary condition }on the Hamiltonian $H_{I}$ in order to arrive to a 
\textit{well defined} KL. Assuming, that the interaction Hamiltonian can be
written (in any particular basis) in the form: $H_{I}=\sum\limits_{\beta
=1}^{n}V_{\beta }\otimes B_{\beta }\,$ with $n\leq N^{2}-1$. The set $%
\left\{ V_{\beta }\right\} _{\beta =1}^{n}$ \textit{must be closed} in the
Heisenberg representation, i.e.: 
\begin{equation}
V_{\beta }(-\tau )=\sum\limits_{\gamma =1}^{m}C_{\beta \gamma }(-\tau
)V_{\gamma }\quad \text{with}\quad m\leq n;  \label{Vii}
\end{equation}%
otherwise the matrix $[a_{\alpha \gamma }]$ will not be positive-definite 
\cite{jpa}. If the KL generator were not a genuine CPS we ought to apply
some random phase approximation (Davies' devices \cite{davies}).

\section{On the Weierstrass shift operators}

Consider the product of two Weierstrass' shift operators, using (\ref{Shift}%
) we write (for $\mathcal{A}>1,b(\mbox{integer})\geq 1$):%
\begin{equation*}
aa^{\dag }=\left( \frac{\mathcal{A-}1}{\mathcal{A}}\right)
^{2}\sum\limits_{n_{1}=0}^{\infty }\ \sum\limits_{l=-\infty }^{\infty }%
\frac{1}{\mathcal{A}^{n_{1}}}\left\vert l-b^{n_{1}}\right\rangle
\left\langle l\right\vert \sum\limits_{n_{2}=0}^{\infty }\
\sum\limits_{l_{4}=-\infty }^{\infty }\frac{1}{\mathcal{A}^{n_{2}}}%
\left\vert l_{4}+b^{n_{2}}\right\rangle \left\langle l_{4}\right\vert .
\end{equation*}%
Then, in Wannier's representation we get the off-diagonal elements%
\begin{eqnarray*}
\left\langle l_{1}\right\vert aa^{\dag }\left\vert l_{2}\right\rangle
&=&\left( \frac{\mathcal{A-}1}{\mathcal{A}}\right)
^{2}\sum\limits_{n_{1},n_{2}=0}^{\infty }\
\sum\limits_{l,l_{3},l_{4}=-\infty }^{\infty }\frac{1}{\mathcal{A}%
^{n_{1}+n_{2}}}\left\langle l_{1}\right. \left\vert l-b^{n_{1}}\right\rangle
\left\langle l\right. \left\vert l_{3}\right\rangle \left\langle
l_{3}\right. \left\vert l_{4}+b^{n_{2}}\right\rangle \left\langle
l_{4}\right. \left\vert l_{2}\right\rangle \\
&=&\left( \frac{\mathcal{A-}1}{\mathcal{A}}\right)
^{2}\sum\limits_{n_{1},n_{2}=0}^{\infty }\ \sum\limits_{l_{3}=-\infty
}^{\infty }\frac{1}{\mathcal{A}^{n_{1}+n_{2}}}\left\langle
l_{1}+b^{n_{1}}\right\vert \left. l_{3}\right\rangle \left\langle
l_{3}\right. \left\vert l_{2}+b^{n_{2}}\right\rangle \\
&=&\left( \frac{\mathcal{A-}1}{\mathcal{A}}\right)
^{2}\sum\limits_{n_{1},n_{2}=0}^{\infty }\frac{\delta
_{l_{1}-l_{2},b^{n_{2}}-b^{n_{1}}}}{\mathcal{A}^{n_{1}+n_{2}}}\quad ,\quad (%
\mathcal{A}>1,b\geq 1).
\end{eqnarray*}%
In a similar way it is simple to show that: 
\begin{equation}
\left\langle l_{1}\right\vert a^{\dag }a\left\vert l_{2}\right\rangle
=\left( \frac{\mathcal{A-}1}{\mathcal{A}}\right)
^{2}\sum\limits_{n_{2},n_{1}=0}^{\infty }\frac{\delta
_{l_{1}-l_{2},b^{n_{1}}-b^{n_{2}}}}{\mathcal{A}^{n_{1}+n_{2}}},\quad (%
\mathcal{A}>1,b\geq 1),  \label{aal1l2}
\end{equation}%
therefore $\left\langle l_{1}\right\vert aa^{\dag }\left\vert
l_{2}\right\rangle =\left\langle l_{1}\right\vert a^{\dag }a\left\vert
l_{2}\right\rangle $, telling that $[a^{\dag },a]=0$. $\qquad \qquad
\blacksquare $

Note that if $b=1$ we explicitly get from (\ref{aal1l2}) that $a^{\dag }a=%
\mathbf{1}$. Assuming that $b=\mbox{integer}$, we can write for the diagonal
elements 
\begin{equation*}
\left\langle l\right\vert a^{\dag }a\left\vert l\right\rangle =\left( \frac{%
\mathcal{A-}1}{\mathcal{A}}\right) ^{2}\sum\limits_{n_{1},n_{2}=0}^{\infty }%
\frac{\delta _{b^{n_{2}},b^{n_{1}}}}{\mathcal{A}^{n_{1}+n_{2}}}=\left\{ 
\begin{array}{c}
\frac{\mathcal{A-}1}{\mathcal{A}+1}<1,\ (\mathcal{A}>1,b>1), \\ 
\\ 
1,\ (\mathcal{A}>1,b=1).%
\end{array}%
\right.
\end{equation*}%
Alternatively, in the Fourier representation we can write:%
\begin{eqnarray}
\left\langle k_{1}\right\vert a^{\dag }a\left\vert k_{2}\right\rangle
&=&\delta \left( k_{1}-k_{2}\right) \left( \frac{\mathcal{A-}1}{\mathcal{A}}%
\right) ^{2}\sum\limits_{n_{2},n_{1}=0}^{\infty }\frac{\exp \left[
-ik_{1}\left( b^{n_{2}}-b^{n_{1}}\right) \right] }{\mathcal{A}^{n_{1}+n_{2}}}
\notag \\
&=&\delta \left( k_{1}-k_{2}\right) \left( \frac{\mathcal{A-}1}{\mathcal{A}}%
\right) ^{2}\sum\limits_{n_{2},n_{1}=0}^{\infty }\frac{\cos k_{1}\left(
b^{n_{2}}-b^{n_{1}}\right) }{\mathcal{A}^{n_{1}+n_{2}}},\quad (\mathcal{A}%
>1,b\geq 1).  \label{aaK}
\end{eqnarray}%
So if $b\neq 1$ (i.e., for the QLW) $a^{\dag }a$ is not the identity
operator, only in the NN case ($b=1$) we get $a^{\dag }a=\mathbf{1}$. From
all these results and considering the structure of the Hamiltonian $H_{S%
\text{ }}$ we may conclude also that $[H_{S\text{ }},a^{\dag }]=[H_{S\text{ }%
},a]=0$, telling us that Weierstrass' shift operators, in the Heisenberg
representation, are constant in time.

Here we calculate the matrix elements of $\left[ \mathbf{q},H_{S}\right] $
for the general case $b\neq 1$. In the Wannier basis we obtain%
\begin{eqnarray}
\left\langle l_{1}\right\vert \left[ \mathbf{q},H_{S}\right] \left\vert
l_{2}\right\rangle &=&\sum_{l}\left\langle l_{1}\right\vert \mathbf{q}%
\left\vert l\right\rangle \left\langle l\right\vert H_{S}\left\vert
l_{2}\right\rangle -\left\langle l_{1}\right\vert H_{S}\left\vert
l\right\rangle \left\langle l\right\vert \mathbf{q}\left\vert
l_{2}\right\rangle  \notag \\
&=&\sum_{l}l\delta _{l_{1},l}\left\langle l\right\vert H_{S}\left\vert
l_{2}\right\rangle -l_{2}\delta _{l,l_{2}}\left\langle l_{1}\right\vert
H_{S}\left\vert l\right\rangle  \notag \\
&=&\left( l_{1}-l_{2}\right) \Omega \left[ \delta _{l_{1},l_{2}}-\frac{%
\mathcal{A-}1}{2\mathcal{A}}\left( \sum_{n}\frac{1}{\mathcal{A}^{n}}\delta
_{l_{1}-l_{2},-b^{n}}+\delta _{l_{1}-l_{2},b^{n}}\right) \right]  \notag \\
&=&\frac{\Omega }{2}\left( \frac{\mathcal{A-}1}{\mathcal{A}}\right) \left(
\sum_{n}\left( \frac{b}{\mathcal{A}}\right) ^{n}\left( \delta
_{l_{1}-l_{2},-b^{n}}-\delta _{l_{1}-l_{2},b^{n}}\right) \right) .
\label{qH}
\end{eqnarray}

Note that when $b=1$ we get $\left[ \mathbf{q},H_{S}\right] =\Omega /2\left(
a-a^{\dag }\right) $, because 
\begin{eqnarray}
\left\langle l_{1}\right\vert \left[ \mathbf{q},H_{S}\right] \left\vert
l_{2}\right\rangle &=&\frac{\Omega }{2}\left( \frac{\mathcal{A-}1}{\mathcal{A%
}}\right) \left( \sum_{n}\left( \frac{1}{\mathcal{A}}\right) ^{n}\left(
\delta _{l_{1}-l_{2},-1}-\delta _{l_{1}-l_{2},1}\right) \right)  \notag \\
&=&\frac{\Omega }{2}\left\langle l_{1}\right\vert \left( a-a^{\dag }\right)
\left\vert l_{2}\right\rangle .  \notag
\end{eqnarray}%
Therefore the operator $\mathbf{p}=\frac{\Omega m}{i2\hbar }\left( a-a^{\dag
}\right) =\frac{\Omega m}{i2\hbar }\left( e^{-\epsilon \partial
_{l}}-e^{\epsilon \partial _{l}}\right) $ can be associated to a
\textquotedblleft discrete\textquotedblright\ momentum operator in a lattice
with scaling parameter $\epsilon $. Also it is trivial to see that $\left[ 
\mathbf{p},H_{S}\right] =0$ as expected for a free particle model. As a
matter of fact, taking the limit of the lattice parameter going to zero,
i.e., 
\begin{eqnarray*}
\lim_{\epsilon \rightarrow 0}\frac{\left( a+a^{\dag }\right) }{2}
&=&\lim_{\epsilon \rightarrow 0}\cosh (\epsilon \partial _{l})\rightarrow (1+%
\frac{1}{2}\epsilon \partial _{l}), \\
\lim_{\epsilon \rightarrow 0}\frac{\left( a-a^{\dag }\right) }{2}
&=&-\lim_{\epsilon \rightarrow 0}\sinh (\epsilon \partial _{l})\rightarrow
-(\epsilon \partial _{l}),
\end{eqnarray*}%
we recover the usual commutation relation $\left[ \mathbf{q},\mathbf{p}%
\right] =-\frac{\Omega m}{i\hbar }\frac{\left( a+a^{\dag }\right) }{2}%
\rightarrow i\hbar \mathbf{1}$ \cite{MOC-CH97}.

\section{The QME for the density matrix of the QLW}

\bigskip

Here we find the QME for our quantum Levy model. To write the QME we have to
calculate the superoperator $F[\bullet ]$ and the effective Hamiltonian $%
H_{eff}$, both objects are given in (\ref{eq4}) and (\ref{eq3})
respectively. We assume that the interaction Hamiltonian (\ref{hi}) is
written in term of two system operators 
\begin{equation}
V_{1}=\hbar \Gamma a=V_{2}^{\dag },\quad \Gamma >0,  \label{Vi}
\end{equation}%
and two baths operators (infinity set of thermal harmonic oscillators \cite%
{haken})%
\begin{equation}
B_{1}=\sum_{k}v_{k}\mathcal{B}_{k}=B_{2}^{\dag },  \label{Bi}
\end{equation}%
which fulfill $\chi _{\alpha \beta }(-\tau )\equiv Tr_{B}\left( \rho
_{B}^{e}\ B_{\alpha }^{\dagger }\,B_{\beta }(-\tau )\right) =\delta _{\alpha
\beta }\chi _{\alpha \alpha }(-\tau )$. So the correlation functions of the
bath are characterized by:%
\begin{eqnarray}
\chi _{1}(-\tau ) &=&\sum_{k}\left\vert v_{k}\right\vert ^{2}\exp \left(
-i\omega _{k}\tau \right) \left( n(\omega )+1\right) ,  \label{cor1} \\
\chi _{2}(-\tau ) &=&\sum_{k}\left\vert v_{k}\right\vert ^{2}\exp \left(
i\omega _{k}\tau \right) n(\omega ),  \label{cor2}
\end{eqnarray}%
where $n(\omega )=\left[ \exp (\hbar \omega /k_{B}T)-1\right] ^{-1}$. Using
that $V_{\alpha }(-\tau )=V_{\alpha }(0),\alpha =1,2$, i.e., they are
constant in time, we can write from (\ref{eq4b}) the fluctuating
superoperator in the form 
\begin{eqnarray}
F\left[ \bullet \right] &=&\left( \frac{\theta }{\hbar }\right)
^{2}\sum\limits_{\alpha }\int_{0}^{\infty }d\tau \left( \chi _{\alpha \alpha
}(-\tau )+\chi _{\alpha \alpha }^{\ast }(-\tau )\right) V_{\alpha }\bullet
V_{\alpha }^{\dagger }  \label{F} \\
&=&\left( \frac{\theta }{\hbar }\right) ^{2}\left( \hbar \Gamma \right)
^{2}\int_{0}^{\infty }d\tau \left[ \left( \chi _{1}(-\tau )+\chi _{1}^{\ast
}(-\tau )\right) a\bullet a^{\dagger }+\left( \chi _{2}(-\tau )+\chi
_{2}^{\ast }(-\tau )\right) a^{\dagger }\bullet a\right] ,  \notag
\end{eqnarray}%
and from (\ref{eq3c}) the effective Hamiltonian as

\begin{eqnarray}
H_{eff}-H_{S} &=&-i\frac{\theta ^{2}}{2\hbar }\sum\limits_{\alpha
}\int_{0}^{\infty }d\tau \left( \chi _{\alpha \alpha }(-\tau )-\chi _{\alpha
\alpha }^{\ast }(-\tau )\right) V_{\alpha }^{\dagger }V_{\alpha }  \label{He}
\\
&=&-i\frac{\theta ^{2}}{2\hbar }\left( \hbar \Gamma \right) ^{2}\left[
\int_{0}^{\infty }d\tau \left( \chi _{1}(-\tau )-\chi _{1}^{\ast }(-\tau
)\right) a^{\dagger }a+\left( \chi _{2}(-\tau )-\chi _{2}^{\ast }(-\tau
)\right) aa^{\dagger }\right] .  \notag
\end{eqnarray}%
From (\ref{cor1}) and (\ref{cor2}) the Fourier transform of the thermal
bath-correlations are%
\begin{eqnarray*}
h_{1}(\omega ) &\equiv &\int_{-\infty }^{+\infty }d\tau \exp (-i\omega )\chi
_{1}(-\tau )=2\pi g(-\omega )\left[ n(-\omega )+1\right] \\
h_{2}(\omega ) &\equiv &\int_{-\infty }^{+\infty }d\tau \exp (-i\omega )\chi
_{2}(-\tau )=2\pi g(\omega )n(\omega ),
\end{eqnarray*}%
where $g(\omega )=\sum_{k}\left\vert v_{k}\right\vert ^{2}\delta \left(
\omega -\omega _{k}\right) $ is the spectral function of the phonon baths 
\cite{caldeira}. In general the half-Fourier transform, that appear in (\ref%
{F}) and \ref{He}), can be written in terms of $h_{\alpha }(\omega )$, i.e.,:%
\begin{equation*}
\int_{0}^{+\infty }d\tau \exp (-i\omega )\chi _{\alpha \alpha }(-\tau )=%
\frac{h_{\alpha }(\omega )}{2}+i\frac{s_{\alpha }(\omega )}{2},
\end{equation*}%
where $s_{\alpha }(\omega )=\frac{1}{\pi }\mathcal{VP}\int_{-\infty
}^{\infty }du\ h_{\alpha }(u)(u-\omega )^{-1}$ is its Hilbert transform.
Note that $h_{\alpha }(\omega )\in \mathcal{R}_{e}$, and so $s_{\alpha
}(\omega )\in \mathcal{R}_{e}$, because the thermal bath correlation
function is stationary $\chi _{\alpha \alpha }(-\tau )=\chi _{\alpha \alpha
}^{\ast }(\tau )$. Then we only need to evaluate the integrals%
\begin{eqnarray*}
\int_{0}^{\infty }d\tau \left( \chi _{\alpha \alpha }(-\tau )+\chi _{\alpha
\alpha }^{\ast }(-\tau )\right) &=&h_{\alpha }(0), \\
\int_{0}^{\infty }d\tau \left( \chi _{\alpha \alpha }(-\tau )-\chi _{\alpha
\alpha }^{\ast }(-\tau )\right) &=&is_{\alpha }(0).
\end{eqnarray*}%
The c-numbers $\{h_{\alpha }(0),s_{\alpha }(0)\}$ are given in terms of the
Fourier representation of the bath correlation function at zero frequency.
Modeling the coupling constant in the Ohmic approximation $g(\omega
)=g\omega $ if $0<\omega <\tilde{\omega}_{c},\ g>0$ (where $\tilde{\omega}%
_{c}$ is larger than any characteristic system's frequency \cite{caldeira})
and taking into account that $\left. \omega n(\omega )\right\vert _{\omega
=0}=\left( \beta \hbar \right) ^{-1}$, we get $h_{1}(\omega \rightarrow
0)=h_{2}(\omega \rightarrow 0)=2\pi g\left( \beta \hbar \right) ^{-1}$ where 
$\beta \equiv 1/k_{B}T$ is the inverse of the temperature of the bath, and $%
s_{1}(\omega \rightarrow 0)=s_{2}(\omega \rightarrow 0)=-2g\tilde{\omega}%
_{c} $. Then%
\begin{eqnarray}
F\left[ \bullet \right] &=&\theta ^{2}\Gamma ^{2}2\pi g\left( \beta \hbar
\right) ^{-1}\left[ a\bullet a^{\dagger }+a^{\dagger }\bullet a\right]
\label{F(a)} \\
&\equiv &\frac{\pi \alpha }{2\hbar \beta }\left[ a\bullet a^{\dagger
}+a^{\dagger }\bullet a\right] ,\quad \alpha >0,  \notag
\end{eqnarray}%
where $\alpha \equiv 4\theta ^{2}\Gamma ^{2}g$ is the dissipative constant.
For the effective Hamiltonian we get 
\begin{eqnarray}
H_{eff} &=&H_{S}-\frac{\theta ^{2}}{2\hbar }\left( \hbar \Gamma \right)
^{2}\left( s_{1}(0)a^{\dagger }a+s_{2}(0)aa^{\dagger }\right)  \notag \\
&=&H_{S}-2g\tilde{\omega}_{c}\theta ^{2}\Gamma ^{2}\hbar a^{\dagger
}a=H_{S}-\hbar \omega _{c}a^{\dagger }a,\quad \omega _{c}>0,  \label{Hef}
\end{eqnarray}%
where $\omega _{c}\equiv 2g\tilde{\omega}_{c}\theta ^{2}\Gamma ^{2}$ is an
upper bound frequency. With these expressions for $F[\bullet ]$ and $H_{eff}$
we can write, using (\ref{eq3c}), (\ref{eq4b}) and (\ref{eq2b}), the QME in
the form%
\begin{eqnarray}
\dot{\rho} &=&\frac{-i}{\hbar }\left[ H_{eff},\rho \right] +F[\rho ]-\frac{1%
}{2}\left\{ F^{\ast }[\mathbf{1}],\rho \right\} _{+}  \label{QMEfull} \\
&=&\frac{-i}{\hbar }\left[ H_{eff},\rho \right] +\frac{\pi \alpha }{4\beta
\hbar }\left( 2a\rho a^{\dag }-a^{\dag }a\rho -\rho a^{\dag }a\right) +\frac{%
\pi \alpha }{4\beta \hbar }\left( 2a^{\dag }\rho a-aa^{\dag }\rho -\rho
aa^{\dag }\right) .  \notag
\end{eqnarray}

Now we explicitly calculate the evolution equation for the elements of the
density matrix $\rho $. Using the Fourier basis in the (\ref{QMEfull}), we
get%
\begin{eqnarray}
\frac{d}{dt}\left\langle k_{1}\right\vert \rho \left\vert k_{2}\right\rangle
&=&\frac{-i}{\hbar }\left\langle k_{1}\right\vert \left[ H_{eff},\rho \right]
\left\vert k_{2}\right\rangle +\frac{\pi \alpha }{2\hbar \beta }\left\langle
k_{1}\right\vert \left( a\rho a^{\dagger }+a^{\dagger }\rho a\right)
\left\vert k_{2}\right\rangle  \label{k1rok2} \\
&&-\frac{\pi \alpha }{2\hbar \beta }\left\langle k_{1}\right\vert \left(
aa^{\dagger }\rho +\rho aa^{\dagger }\right) \left\vert k_{2}\right\rangle ,
\notag
\end{eqnarray}%
where $H_{eff}=H_{S}-\hbar \omega _{c}a^{\dag }a$, and we have used that $%
\left[ a,a^{\dagger }\right] =0$. Note that%
\begin{eqnarray}
\left\langle k_{1}\right\vert aa^{\dagger }\rho \left\vert
k_{2}\right\rangle &=&\int_{-\pi }^{\pi }dk\left\langle k_{1}\right\vert
aa^{\dagger }\left\vert k\right\rangle \left\langle k\right\vert \rho
\left\vert k_{2}\right\rangle  \label{aaro} \\
&=&\int_{-\pi }^{\pi }dk\delta \left( k_{1}-k\right) \left( \frac{\mathcal{A-%
}1}{\mathcal{A}}\right) ^{2}\sum\limits_{n_{2},n_{1}=0}^{\infty }\frac{\cos
k_{1}\left( b^{n_{2}}-b^{n_{1}}\right) }{\mathcal{A}^{n_{1}+n_{2}}}%
\left\langle k\right\vert \rho \left\vert k_{2}\right\rangle  \notag \\
&=&\left( \frac{\mathcal{A-}1}{\mathcal{A}}\right)
^{2}\sum\limits_{n_{2},n_{1}=0}^{\infty }\frac{\cos k_{1}\left(
b^{n_{2}}-b^{n_{1}}\right) }{\mathcal{A}^{n_{1}+n_{2}}}\left\langle
k_{1}\right\vert \rho \left\vert k_{2}\right\rangle ,  \notag
\end{eqnarray}%
and%
\begin{eqnarray}
\left\langle k_{1}\right\vert \rho aa^{\dagger }\left\vert
k_{2}\right\rangle &=&\int_{-\pi }^{\pi }dk\left\langle k_{1}\right\vert
\rho \left\vert k\right\rangle \delta \left( k_{2}-k\right) \left( \frac{%
\mathcal{A-}1}{\mathcal{A}}\right) ^{2}\sum\limits_{n_{2},n_{1}=0}^{\infty }%
\frac{\cos k_{2}\left( b^{n_{2}}-b^{n_{1}}\right) }{\mathcal{A}^{n_{1}+n_{2}}%
}  \label{roaa} \\
&=&\left( \frac{\mathcal{A-}1}{\mathcal{A}}\right)
^{2}\sum\limits_{n_{2},n_{1}=0}^{\infty }\frac{\cos k_{2}\left(
b^{n_{2}}-b^{n_{1}}\right) }{\mathcal{A}^{n_{1}+n_{2}}}\left\langle
k_{1}\right\vert \rho \left\vert k_{2}\right\rangle ,  \notag
\end{eqnarray}%
where we have used (\ref{aaK}). Using that%
\begin{equation*}
\left\langle k_{1}\right\vert a\left\vert k_{2}\right\rangle =\left( \frac{%
\mathcal{A-}1}{\mathcal{A}}\right) \sum\limits_{n}^{\infty }\frac{\exp
ik_{2}b^{n}}{\mathcal{A}^{n}}\delta \left( k_{1}-k_{2}\right) ,
\end{equation*}%
we can also write 
\begin{eqnarray}
\left\langle k_{1}\right\vert a\rho a^{\dagger }\left\vert
k_{2}\right\rangle &=&\int_{-\pi }^{\pi }dk\int_{-\pi }^{\pi
}dk_{3}\left\langle k_{1}\right\vert a\left\vert k\right\rangle \left\langle
k\right\vert \rho \left\vert k_{3}\right\rangle \left\langle
k_{3}\right\vert a^{\dagger }\left\vert k_{2}\right\rangle  \label{ara1} \\
&=&\left( \frac{\mathcal{A-}1}{\mathcal{A}}\right)
^{2}\sum\limits_{n,m}^{\infty }\frac{\exp ik_{1}b^{n}}{\mathcal{A}^{n}}%
\frac{\exp -ik_{2}b^{m}}{\mathcal{A}^{m}}\left\langle k_{1}\right\vert \rho
\left\vert k_{2}\right\rangle ,  \notag
\end{eqnarray}%
and%
\begin{equation}
\left\langle k_{1}\right\vert a^{\dagger }\rho a\left\vert
k_{2}\right\rangle =\left( \frac{\mathcal{A-}1}{\mathcal{A}}\right)
^{2}\sum\limits_{n,m}^{\infty }\frac{\exp -ik_{1}b^{n}}{\mathcal{A}^{n}}%
\frac{\exp ik_{2}b^{m}}{\mathcal{A}^{m}}\left\langle k_{1}\right\vert \rho
\left\vert k_{2}\right\rangle .  \label{ara11}
\end{equation}%
Therefore putting (\ref{aaro},\ref{roaa},\ref{ara1},\ref{ara11}) in (\ref%
{k1rok2}) we get the final result (for $b\geq 1,\mathcal{A}>1$)%
\begin{eqnarray}
\frac{d}{dt}\left\langle k_{1}\right\vert \rho \left\vert k_{2}\right\rangle
&=&\frac{-i}{\hbar }\left( \mathrm{E}_{k_{1}}-\mathrm{E}_{k_{2}}\right)
\left\langle k_{1}\right\vert \rho \left\vert k_{2}\right\rangle
\label{k1rok22} \\
&&+\frac{\pi \alpha }{2\hbar \beta }\left( \frac{\mathcal{A-}1}{\mathcal{A}}%
\right) ^{2}\sum\limits_{n,m=0}^{\infty }\frac{1}{\mathcal{A}^{n+m}}\left[
2\cos \left( k_{1}b^{n}-k_{2}b^{m}\right) \right.  \notag \\
&&-\left. \cos k_{1}\left( b^{n}-b^{m}\right) -\cos k_{2}\left(
b^{n}-b^{m}\right) \right] \left\langle k_{1}\right\vert \rho \left\vert
k_{2}\right\rangle ,  \notag
\end{eqnarray}%
where $\mathrm{E}_{k_{1}}$ is the effective eigenenergy:%
\begin{eqnarray}
\mathrm{E}_{k_{1}} &\equiv &\Omega \left( 1-\frac{\mathcal{A-}1}{\mathcal{A}}%
\sum_{n=0}^{\infty }\frac{1}{\mathcal{A}^{n}}\cos \left( b^{n}k_{1}\right)
\right)  \notag \\
&&-\hbar \omega _{c}\left( \frac{\mathcal{A-}1}{\mathcal{A}}\right)
^{2}\sum\limits_{n_{2},n_{1}=0}^{\infty }\frac{\cos k_{1}\left(
b^{n_{2}}-b^{n_{1}}\right) }{\mathcal{A}^{n_{1}+n_{2}}}.
\end{eqnarray}

\section{The second moment of the QLW}

\bigskip

In the present appendix we explicitly calculate the quantum thermal
mean-value $\left\langle \mathbf{q}^{2}(t)\right\rangle $, thus%
\begin{equation}
\left\langle \mathbf{q}^{2}(t)\right\rangle =Tr\left[ \mathbf{q}^{2}\rho (t)%
\right] =\sum_{l=-\infty }^{\infty }\int_{-\pi }^{\pi }\int_{-\pi }^{\pi
}dk_{1}dk_{2}\left\langle l\right\vert \mathbf{q}^{2}\left\vert
k_{1}\right\rangle \left\langle k_{1}\right\vert \rho (t)\left\vert
k_{2}\right\rangle \left\langle k_{2}\right. \left\vert l\right\rangle ,
\label{Luno}
\end{equation}%
where $\left\langle k_{2}\right\vert \rho (t)\left\vert k_{1}\right\rangle $
is known form our general solution Eq.(\ref{evo2}). Using the pure state (%
\ref{pure}) as the initial condition for the density matrix we get%
\begin{eqnarray}
\left\langle \mathbf{q}^{2}(t)\right\rangle &=&\sum_{l,l_{1}=-\infty
}^{\infty }\int_{-\pi }^{\pi }dk_{1}\int_{-\pi }^{\pi }dk_{2}\left\langle
l\right\vert \mathbf{q}^{2}\left\vert l_{1}\right\rangle \left\langle
l_{1}\right. \left\vert k_{1}\right\rangle \left\langle k_{1}\right\vert
\rho (t)\left\vert k_{2}\right\rangle \left\langle k_{2}\right. \left\vert
l\right\rangle  \label{qq} \\
&=&\frac{1}{2\pi }\sum_{l,l_{2}=-\infty }^{\infty }\int_{-\pi }^{\pi
}\int_{-\pi }^{\pi }dk_{1}dk_{2}\ l^{2}\delta
_{l,l_{1}}e^{ik_{1}l_{1}}\left\langle k_{1}\right\vert \rho (t)\left\vert
k_{2}\right\rangle e^{-ik_{2}l}  \notag \\
&=&\frac{1}{(2\pi )^{2}}\sum_{l=-\infty }^{\infty }\int_{-\pi }^{\pi
}\int_{-\pi }^{\pi }dk_{1}dk_{2}\ l^{2}e^{i(k_{1}-k_{2})(l-l_{0})}\ \exp
\left( \mathcal{F}\left( k_{1},k_{2},b,\mathcal{A}\right) t\right) .  \notag
\end{eqnarray}%
Introducing the change of variable $l-l_{0}\rightarrow l$, using the
"functional" properties of the Dirac-delta \cite{Olaf}%
\begin{eqnarray*}
\frac{1}{2\pi }\sum\limits_{l=-\infty }^{\infty }\left( il\right)
^{n}e^{i(k_{1}-k_{2})l} &=&(-1)^{n}\delta ^{(n)}\left( k_{1}-k_{2}\right) ,
\\
\int f(k)\delta ^{(n)}\left( k-k_{0}\right) dk &=&\left( -1\right)
^{n}f^{(n)}\left( k_{0}\right) ,
\end{eqnarray*}%
and after integrating by part we finally get the result

\begin{eqnarray*}
\left\langle \mathbf{q}^{2}(t)\right\rangle &=&\frac{1}{(2\pi )^{2}}%
\sum_{l=-\infty }^{\infty }\int_{-\pi }^{\pi }\int_{-\pi }^{\pi
}dk_{1}dk_{2}\ \left( l+l_{0}\right) ^{2}e^{i(k_{1}-k_{2})l}\ \exp \left( 
\mathcal{F}\left( k_{1},k_{2},b,\mathcal{A}\right) t\right) \\
&=&-\frac{1}{2\pi }\int_{-\pi }^{\pi }dk_{2}\left\{ \left[ \frac{d^{2}%
\mathcal{F}\left( k,k_{2},b,\mathcal{A}\right) }{dk^{2}}t\right] +\left[ 
\frac{d\mathcal{F}\left( k,k_{2},b,\mathcal{A}\right) }{dk}t\right]
^{2}\right\} _{k=k_{2}} \\
&&-\frac{l_{0}}{i\pi }\int_{-\pi }^{\pi }dk_{2}\left\{ \frac{d\mathcal{F}%
\left( k,k_{2},b,\mathcal{A}\right) }{dk}t\right\} _{k=k_{2}}+\frac{l_{0}^{2}%
}{2\pi }\int_{-\pi }^{\pi }dk_{2},
\end{eqnarray*}%
where we have used that $\exp \left( \mathcal{F}\left( k_{2},k_{2},b,%
\mathcal{A}\right) t\right) =1$. By introducing the explicit expression (\ref%
{evo3}), for $\mathcal{F}\left( k,k_{2},b,\mathcal{A}\right) ,$ and noting
that we have assumed $b=\mbox{integer}$, we finally obtain (\ref{q2det}).

As we have already pointed out when working on the differential equation for
the second moment, see (\ref{q2}) and (\ref{dqdt2}), here we have explicitly
shown that\ only if $b<\mathcal{A}$ we get $\left\langle \mathbf{q}%
^{2}(t)\right\rangle \neq \infty $, this conclusion is quite different when
compared to the \textit{classical} counterpart of a Levy flight, where a
divergent second moment appears if $b^{2}>\mathcal{A}$. The reason is that
for the QLW the second moment $\left\langle \mathbf{q}^{2}(t)\right\rangle $
cannot just be calculated from the second derivative of the Fourier
transform of the space-probability distribution \cite{kampen,libro}. In
quantum mechanics the thermal second moment is given through a trace
operation, and $\left\langle k_{1}\right\vert \rho (t)\left\vert
k_{2}\right\rangle $ takes into account all the coherence phenomena.

\section{Density of relaxation for a classical homogeneous walk and its
localized probability}

\bigskip

Consider a classical one dimensional homogeneous random walk, its master
equation will be%
\begin{equation*}
\frac{d}{dt}P\left( s,t\right\vert \left. 0,0\right) =D\left[ P\left(
s+1,t\right\vert \left. 0,0\right) +P\left( s-1,t\right\vert \left.
0,0\right) -2P\left( s,t\right\vert \left. 0,0\right) \right] ,\quad P\left(
s,0\right\vert \left. 0,0\right) =\delta _{s,0}.
\end{equation*}%
Introducing the Fourier transform of the probability $\lambda \left(
k,t\right) =\sum_{s=-\infty }^{\infty }e^{iks}P\left( s,t\right\vert \left.
0,0\right) $, we get the solution $\lambda \left( k,t\right) =\exp \left( 2Dt%
\left[ \cos k-1\right] \right) .$ Therefore the relaxation form the initial
condition (classical localized probability function) will be%
\begin{equation}
P_{0}(t)\equiv \frac{1}{2\pi }\int_{-\pi }^{\pi }\lambda \left( k,t\right)
dk=\int_{0}^{4D}e^{-\gamma t}\rho (\gamma )d\gamma ;\quad \gamma \geq 0.
\label{Pt}
\end{equation}%
Where $\rho (\gamma )$ is the (probability of) relaxation density:%
\begin{equation*}
\rho (\gamma )=\frac{1}{2\pi }\int_{-\pi }^{\pi }\delta \left( \gamma -2D%
\left[ 1-\cos k\right] \right) dk=\frac{1}{\pi }\left( 4\gamma D-\gamma
^{2}\right) ^{-1/2},
\end{equation*}%
in total analogy to the density of states of the 1D \textit{tight-binding}
model \cite{libro,Economou}. From the expression $\rho (\gamma )$ and Eq. (%
\ref{Pt}) it is simple, by using the Laplace transform, to show that
asymptotically for $t\rightarrow \infty $%
\begin{equation}
P_{0}(t)\sim \frac{1}{\sqrt{t}},  \label{Pt2}
\end{equation}%
this is the typical behavior for the 1D localized probability in a classical
diffusive regime.\newpage

%figure Captions:

\end{document}